\documentstyle[aps,pra,preprint,psfig]{revtex}

\begin{document}

\vspace*{5cm}
\begin{center}
{\large\bf PSEUDOGAPS AND MAGNETIC PROPERTIES\\
OF THE TWO-DIMENSIONAL\mbox{\boldmath$t$-$J$} MODEL

}

\vspace{3ex}
A.~Sherman$^a$ and M.~Schreiber$^b$

\vspace{2ex}
$^a$Institute of Physics, University of Tartu, Riia 142,
EE-2400 Tartu, Estonia

$^b$Institut f\"ur Physik, Technische Universit\"at,
D-09107 Chemnitz, Germany
\end{center}

\section{Introduction}
The photoemission and magnetic properties of cuprate perovskites have
been extensively studied during the last few years, both because of
their unusual behaviour and in the hope that they might provide
insight into the physical origin of high-temperature
superconductivity.  Among these properties the pseudogap observed in
photoemission \cite{loeser,marshall,ding} and the magnetic pseudogap
revealed in the static susceptibility and in the spin-lattice
relaxation rate of normal-state underdoped cuprates
\cite{imai89,rossat} have attracted considerable attention.  A number
of different approaches were suggested for the description of the
pseudogaps.  In particular, in view of the similarity in symmetry and
size of the photoemission pseudogap with the superconducting gap, in
Refs.~\cite{ding,maly,markiewicz,kristoffel} this pseudogap was
connected with the superconducting fluctuations existing above $T_c$.
This idea was based on earlier theoretical results of
Refs.~\cite{trivedi,emery95}.  Another point of view was suggested in
Ref.~\cite{sher97b} where the energy spectrum of the two-dimensional
(2D) $t$-$J$ model was shown to have a peculiarity which is similar by
its properties to the photoemission pseudogap.  In accord with
Ref.~\cite{sher97b} the pseudogap is a consequence of a specific
dispersion of the strongly correlated electron system at moderate
doping and is not connected with superconducting fluctuations.  The
discussion of the magnetic pseudogap in doped cuprates is mainly based
on scaling arguments \cite{sachdev,sokol,barzykin} and on the idea of
real-space pairing \cite{loktev}.

In this paper we describe the energy spectrum, including the
photoemission pseudogap, and the magnetic properties demonstrating the
magnetic pseudogap in a unified approach based on the 2D $t$-$J$ model
widely used for the description of CuO$_2$ planes of cuprate
perovskites (the extensive literature on this model is reviewed in
Ref.~\cite{izyumov}).  For the consideration of the paramagnetic state
we extend the spin-wave theory with the constraint of zero staggered
magnetization, developed for the Heisenberg model
\cite{takahashi,tang}, to the $t$-$J$ model.  The spectrum is
determined by solving numerically a set of self-energy equations for
hole and magnon Green's functions in the self-consistent Born
approximation with account of the constraint \cite{sher98}.  The
constraint can be fulfilled in the ranges of hole concentrations
$0.02\lesssim x\lesssim 0.17$ and temperatures $T\lesssim 100$~K\@.
In this region the obtained hole spectrum differs from a conventional
metallic spectrum as the quasiparticle weights of states are less than
1 and change with the hole concentration.  This leads to the violation
of Luttinger's theorem.

The hole spectrum consists of two essentially different parts:  a
persistent portion of the narrow spin-polaron band, which is typical
for the low-concentration ($x\lesssim 0.02$) spectrum, and a wider
part appearing from $x\approx 0.04$ which is characterized by the
energy parameter $t$, the hopping constant.  The former part provides
the most intensive features in the hole spectral function near the
Fermi level.  For $x<0.17$ this part, which is pinned to the Fermi
level near $(\pm\pi/2,\pm\pi/2)$, bends upwards (in the hole picture)
on approaching $(\pm\pi,0)$, $(0,\pm\pi)$.  In the hole spectral
function the crossing of the Fermi level by the second, wider part is
completely lost within the foot of a more intensive spin-polaron peak
in this region of the Brillouin zone.  This looks like the
disappearance of a part of the Fermi surface and the opening of a
pseudogap near $(\pm\pi,0)$, $(0,\pm\pi)$.  Obtained size, symmetry
and concentration dependence of the pseudogap are in agreement with
photoemission data in $\rm Bi_2Sr_2CaCu_2O_{8+\delta}$ (Bi2212)
\cite{loeser,marshall,ding}.

Another peculiarity of the calculated hole spectrum is an extended
saddle point near $(0,\pi)$.  By the energy position and by the
extension in the Brillouin zone our results reproduce well the
analogous feature of the photoemission spectra.  In the case of
optimal doping, $x\approx 0.17$, this peculiarity leads to a 2D Fermi
surface in the considered 2D system.

A gap in the magnon spectrum of the undoped antiferromagnet appears as
a consequence of the constraint of zero staggered magnetization
\cite{takahashi,tang}.  Starting from $x\approx 0.02$ this gap is
filled by overdamped magnons.  Their increased damping is the
consequence of the hole-magnon interaction and at $T=0$ it indicates
the destruction of the long-range antiferromagnetic order by holes
\cite{sher93a}.  The arising pseudogap leads to the decrease of the
static spin susceptibility and the spin-lattice relaxation rate
$(T_1T)^{-1}$ with decreasing temperature, as observed
experimentally.  This behavior is typical for the quantum disordered
regime.  Calculated values of these quantities are in qualitative and
in some cases in quantitative agreement with experiment in underdoped
YBa$_2$Cu$_3$O$_{6+\delta}$.  In some articles the photoemission and
magnetic pseudogaps are identified (see, e.g., Ref.~\cite{ding}).  In
our opinion these pseudogaps are two different, unconnected
peculiarities of the spectra of two well-distinguishable
subsystems --- holes and magnons.

The outline of the paper is as follows.  In Sec.~II we discuss the
derivation of the $t$-$J$ Hamiltonian from the extended Hubbard
Hamiltonian which is supposed to give a realistic description of
CuO$_2$ planes.  This allows us to rewrite spin operators of oxygen
and copper ions in terms of operators of the $t$-$J$ model.  The
extension of the modified spin-wave approximation with zero staggered
magnetization on the $t$-$J$ model is discussed in Sec.~III\@.  The
self-energy equations for the hole and magnon Green's functions and
the hole and magnon contributions to the magnetic susceptibility in
terms of these functions are written out in Sec.~IV\@.  Numerical
results on the spectrum are considered in Sec.~V\@.  The pseudogap in
the hole spectrum and magnetic properties are discussed in Sec.~VI and
VII, respectively.  Finally our conclusions are given in Sec.~VIII.

\section{The effective Hamiltonian}
The extended Hubbard model \cite{emery87} is widely used for the
description of CuO$_2$ planes of cuprate perovskites.  The
Hamiltonian of the model can be written in the form \cite{sher93}
\begin{eqnarray}
&&H=\sum_{\bf m}H_{\bf m}+2t_{pd}\lambda_{\bf a}\sum_{\bf ma\sigma}
 (d_{\bf m\sigma}^\dagger\phi_{\bf m+a,\sigma}+{\rm H.c.}),
 \nonumber \\[-1.2ex]
&&        \label{ehh}\\[-1.2ex]
&&H_{\bf m}=Un_{{\bf m},+1}n_{{\bf m},-1}+\Delta\sum_\sigma
 \phi^\dagger_{\bf m\sigma}\phi_{\bf m\sigma}
 +2t_{pd}\lambda_{\bf 0}
 \sum_\sigma(d_{\bf m\sigma}^\dagger\phi_{\bf m\sigma}+{\rm H.c.}),
\nonumber
\end{eqnarray}
where $d^\dagger_{\bf m\sigma}$ is the creation operator of electrons
in the $3d_{x^2-y^2}$ orbitals of copper at the plane site {\bf m}
with the spin $\sigma=\pm 1$, $\phi^\dagger_{\bf m\sigma}$ is the
Fourier transform of the operator $\phi^\dagger_{\bf
k\sigma}=(\beta_{\bf k}/2\sqrt{N}) \sum_{\bf ma}\exp(-i{\bf km})
p^\dagger_{{\bf m+a}/2,\sigma}$ constructed from the creation
operators of electrons in the $2p_\sigma$ orbitals of oxygen
$p^\dagger_{{\bf m+a}/2,\sigma}$.  Complementary linear combinations
of these operators, which do not hybridize with the $3d_{x^2-y^2}$
copper orbitals, are omitted in Eq.~(\ref{ehh}) because their energy
is much higher.  In Eq.~(\ref{ehh}), ${\bf a}=(\pm a,0),(0,\pm a)$
where $a$ is the in-plane copper distance which is taken as the unit
of length, $\beta_{\bf k}=\{1+[\cos(k_x)+\cos(k_y)]/2\}^{-1/2}$, $N$
is the number of sites; $U$, $\Delta$, and $t_{pd}$ are the Hubbard
repulsion on copper, the Cu-O promotion energy and hybridization,
respectively, $n_{\bf m\sigma}=d^\dagger_{\bf m\sigma}d_{\bf
m\sigma}$, $\lambda_{\bf m} = N^{-1}\sum_{\bf k}\exp(i{\bf km})
\beta_{\bf k}^{-1}$, $\lambda_{\bf 0}\approx 0.96$, $\lambda_{\bf
a}\approx 0.14$.  Other components of $\lambda_{\bf m}$ are small and
the respective terms are omitted in Eq.~(\ref{ehh}).

The splitting of the Hamiltonian into the one- and two-site parts in
Eq.~(\ref{ehh}) provides a good starting point for the perturbation
theory, because for parameters \cite{mcmahan} of La$_2$CuO$_4$ these
two parts are characterized by energies differing by one order of
magnitude.  Notice that the frequently used perturbation expansion in
powers of $t_{pd}$ does not work for these parameters, as the
hybridization is actually not small in comparison with other energies
\cite{sher93,jefferson}.  The prefactor $\lambda_{\bf a}\sim 0.1$ in
Eq.~(\ref{ehh}) allows one to overcome this difficulty.  The
zero-order, one-site part $\sum_{\bf m}H_{\bf m}$ of the Hamiltonian
has two sets of states corresponding to unoccupied and occupied site
states of the $t$-$J$ model
\begin{eqnarray}
&&\vert{\bf m}\rangle=\biggl[\frac{c_{21}}{\sqrt{2}}
 (\phi^\dagger_{{\bf m},+1}d^\dagger_{{\bf m},-1}-
 \phi^\dagger_{{\bf m},-1}d^\dagger_{{\bf m},+1})
 +c_{22}\phi_{{\bf m},-1}^\dagger\phi_{{\bf m},+1}^\dagger
 +c_{23}d_{{\bf m},-1}^\dagger d_{{\bf
 m},+1}^\dagger\biggr]\vert v_{\bf m}\rangle,\nonumber\\[-1.2ex]
&&         \label{es}\\[-1.2ex]
&&\vert{\bf m}\sigma\rangle=(c_{31}\phi^\dagger_{{\bf m},-1}
 \phi^\dagger_{{\bf m},+1}d^\dagger_{\bf m\sigma}
 +c_{32}\phi^\dagger_{\bf m\sigma}d_{{\bf m},-1}^\dagger
 d_{{\bf m},+1}^\dagger)\vert v_{\bf m}\rangle,\nonumber
\end{eqnarray}
where $\vert v_{\bf m}\rangle$ is the site vacuum state and the
coefficients $c_{ij}$ are obtained in the course of the
diagonalization of $H_{\bf m}$.  For a given number of holes crystal
states $\vert q\rangle$ constructed as products of site states
(\ref{es}) form the degenerate ground states of the zero-order
Hamiltonian $H_0=\sum_{\bf m}H_{\bf m}$.  These ground states are
separated by a finite gap of the order of $\min(\Delta,t_{pd})$ from
excited states.  In such conditions one can use the operator form of
the perturbation theory \cite{tyablikov} to obtain an effective
Hamiltonian acting in the subspace of the low-lying states $\vert
q\rangle$.  Up to the terms of the second order in the perturbation
$H_1$ [the two-site part of Hamiltonian (\ref{ehh})] this effective
Hamiltonian reads
$$H_{\rm eff}={\cal P}[H_1-H_1(1-{\cal P})(H_0-E_0)^{-1}(1-{\cal P})
H_1]{\cal P},$$
where ${\cal P}=\sum_q\vert q\rangle\langle q\vert$ and $E_0$ is the
eigenenergy of $H_0$ for the ground states $\vert q\rangle$.  Using
Eqs.~(\ref{ehh}) and (\ref{es}) we obtain the effective $t$-$J$
Hamiltonian \cite{sher93}
\begin{equation}
H_{\rm eff}=t\sum_{\bf ma\sigma}\vert{\bf m+a},\sigma\rangle
 \langle{\bf m+a}\vert \vert{\bf m}\rangle\langle{\bf m}\sigma\vert
 \mbox{}+\frac{J}{2}\sum_{\bf ma}{\bf S_mS}_{\bf m+a},
\label{tjh}\end{equation}
where $S_{\bf m}^\sigma=S_{\bf m}^x+i\sigma S_{\bf m}^y= \vert{\bf
m}\sigma\rangle\langle{\bf m},-\sigma\vert$, $S_{\bf
m}^z=\sum_\sigma(\sigma/2)\vert{\bf m}\sigma\rangle\langle {\bf
m}\sigma\vert$ are the components of the spin operator ${\bf S_m}$,
$t$ and $J$ are the effective hopping and superexchange constants
which are expressed in terms of the parameters of the extended Hubbard
Hamiltonian [see Ref.~\cite{sher93}; in Eq.~(\ref{tjh}), we omitted
three-site ($t'$) terms which give only small corrections in the
considered case $J/t\ll 1$].  Using parameters of Ref.~\cite{mcmahan}
we estimated the ratio $J/t$ to lie in the range 0.1--0.5.  In the
following discussion we use $J/t=0.2$ and $t=0.5$~eV.

The Zeeman term of the Hamiltonian can be written in the form
\begin{equation}
H_Z=2\mu_B\sum_{\bf m}{\bf s_mH_m}+\mu_B\sum_{\bf ma}{\bf s}_{{\bf
m+a}/2}{\bf H}_{{\bf m+a}/2},
\label{zt}\end{equation}
where $\mu_B$ is the Bohr magneton, ${\bf H_m}$ is the applied
magnetic field, and $\bf s_m$ and $\bf s_{{\bf m+a}/2}$ are composed of
\begin{eqnarray*}
&&s^\sigma_{\bf m}=d^\dagger_{\bf m\sigma}d_{\bf m,-\sigma},\quad
 s^z_{\bf m}=\sum_\sigma\frac{\sigma}{2}d^\dagger_{\bf m\sigma}
 d_{\bf m\sigma},\\[-0.5ex]
&&s^\sigma_{{\bf m+a}/2}=p^\dagger_{{\bf m+a}/2,\sigma}p_{{\bf
 m+a}/2,-\sigma},\quad
s^z_{{\bf m+a}/2}=\sum_\sigma\frac{\sigma}{2}p^\dagger_{{\bf
 m+a}/2,\sigma}p_{{\bf m+a}/2,\sigma},
\end{eqnarray*}
respectively.  Using notations of Eq.~(\ref{ehh}) the oxygen spin
operators can be approximately rewritten as
\begin{eqnarray}
s^\sigma_{{\bf m+a}/2}&=&\frac{\tilde{\beta}^2}{4}(\phi^\dagger_{\bf
 m\sigma}+\phi^\dagger_{\bf m+a,\sigma})(\phi_{\bf m,-\sigma}+
 \phi_{\bf m+a,-\sigma}),\nonumber\\[-1.7ex]
&&\label{oso}\\[-1.7ex]
s^z_{{\bf m+a}/2}&=&\frac{\tilde{\beta}^2}{8}\sum_\sigma\sigma
 (\phi^\dagger_{\bf m\sigma}+\phi^\dagger_{\bf m+a,\sigma})
 (\phi_{\bf m\sigma}+\phi_{\bf m+a,\sigma}),\nonumber
\end{eqnarray}
where $\tilde{\beta}=\beta_{\bf 0}+\beta_{\bf a}\approx 0.96$,
$\beta_{\bf m}$ being the Fourier transform of $\beta_{\bf k}$.  In the
basis of states (\ref{es}) the spin operators read
\begin{eqnarray}
&&s^z_{{\bf m+a}/2}=c^2_{21}c^2_{31}\tilde{\beta}^2
 \sum_\sigma\frac{\sigma}{16}(\vert{\bf m}\sigma\rangle\langle{\bf m}|
 \vert{\bf m+a}\rangle\langle{\bf m+a},\sigma| \nonumber\\
&&\quad\quad\quad\quad +\vert{\bf m+a},\sigma\rangle\langle
 {\bf m+a}|\vert{\bf m}\rangle\langle{\bf m}\sigma|),\label{zso}\\[1ex]
&&s^z_{\bf m}=c^2_{31}\sum_\sigma\frac{\sigma}{2}\vert{\bf m}\sigma
 \rangle\langle{\bf m}\sigma|,\nonumber
\end{eqnarray}
where in the oxygen operator terms containing small coefficients
$c_{32}$, $c_{23}$ were omitted.  In the following discussion we set
$c_{21}=c_{31}=\tilde{\beta}=1$.

\section{The spin-wave approximation\newline
for zero staggered magnetization}
Hamiltonian (\ref{tjh}) can be essentially simplified with the use of
the spin-wave approximation.  As known \cite{birgeneau}, at low
temperatures and hole concentrations $x\lesssim 0.02$ the CuO$_2$
planes are antiferromagnetically ordered.  For larger $x$ this
long-range ordering is destroyed.  To describe low-lying magnetic
excitations and their interaction with holes in this case we use the
version of the spin-wave theory formulated in
Refs.~\cite{takahashi,tang} for the Heisenberg antiferromagnet with
zero staggered magnetization.  As shown in
Refs.~\cite{takahashi,tang}, in the absence of holes this approach
reproduces results obtained in Refs.~\cite{arovas,chakravarty} with
the mean-field Schwinger boson and renormalization group theories and
it is remarkably accurate, as follows from the comparison with exact
diagonalization and Monte Carlo results.  Here we reformulate this
approach to simplify the inclusion of holes in the Hamiltonian.

We use the Holstein-Primakoff transformation \cite{tyablikov} to
introduce boson operators of spin waves $b_{\bf m}$,
\begin{equation}
S^z_{\bf m}={\rm e}^{i\bf\Pi m}\left(\frac{1}{2}-
 b^\dagger_{\bf m}b_{\bf m}\right),\quad
S^\sigma_{\bf m}=P^\sigma_{\bf m}\varphi_{\bf m}b_{\bf m}+
 P^{-\sigma}_{\bf m}b^\dagger_{\bf m}\varphi_{\bf m},
\label{swa}\end{equation}
where ${\bf\Pi}=(\pi,\pi)$, $P^\sigma_{\bf m}=[1+\sigma\exp(i{\bf\Pi
m})]/2$, and $\varphi_{\bf m}=(1-b^\dagger_{\bf m}b_{\bf m})^{1/2}$.
In Eq.~(\ref{swa}), the factors $\exp(i{\bf\Pi m})$ and
$P^{\pm\sigma}_{\bf m}$ are introduced to account for alternating
directions of spins in the classical N\'eel state which is used as the
reference state in the spin-wave approximation.  On substituting
Eq.~(\ref{swa}) into the Heisenberg part of Hamiltonian (\ref{tjh}),
expanding $\varphi_{\bf m}$ and keeping terms up to the quartic order,
we use the mean-field approximation in these latter terms
\begin{equation}
H_H=\frac{J}{2}\sum_{\bf ma}{\bf S_{m+a}S_m}
 \approx-\frac{JN}{2}-J\langle b_{\bf 0}b_{\bf a}\rangle
 \biggl[
 4\sum_{\bf m}b^\dagger_{\bf m}b_{\bf m}+\frac{1}{2}\sum_{\bf ma}
 \left(b^\dagger_{\bf m+a}b^\dagger_{\bf m}+b_{\bf m+a}b_{\bf m}
 \right)\biggr],
\label{hh}\end{equation}
where angular brackets denote averaging over the grand canonical
ensemble and the four correlations $\langle b_{\bf 0}b_{\bf a}\rangle$
are supposed to be equal.  On deriving Eq.~(\ref{hh}) we took into
account the condition
\begin{mathletters}
\label{zsm}
\begin{equation}
\langle b^\dagger_{\bf m}b_{\bf m}\rangle=\frac{1}{2}
\end{equation}
which follows from the constraint of zero staggered magnetization,
\begin{equation}
\sum_{\bf m}{\rm e}^{i\bf\Pi m}S^z_{\bf m}=0\quad {\rm or}\quad
 \sum_{\bf m}b^\dagger_{\bf m}b_{\bf m}=\frac{N}{2}
\end{equation}
\end{mathletters}
and ensures zero site magnetization, $\langle S^z_{\bf m}\rangle=0$.
To account for this constraint in the subsequent consideration we add
the term $2J\nu\langle b_{\bf 0}b_{\bf a}\rangle \sum_{\bf
m}b^\dagger_{\bf m}b_{\bf m}$ with the Lagrange multiplier $\nu$ to
Hamiltonian (\ref{hh}).

The deviation of the quartic terms from their mean-field value in
Eq.~(\ref{hh}) describes the magnon-magnon interaction which leads to
the magnon damping \cite{tyc}.  In the $t$-$J$ model there is another
mechanism of the magnon damping connected with the hole-magnon
interaction.  Estimations \cite{sher93a,tyc} show that in the
considered range of hole concentrations this latter interaction gives
the main contribution to the damping.  Therefore we do not consider
the magnon-magnon interaction below.

The resulting Hamiltonian is diagonalized by the unitary transformation
\begin{equation}
U=\exp\biggl[\frac{1}{2}\sum_{\bf k\sigma}\!
 \stackrel{\vphantom{a}}{\vphantom{a}}^\prime\!\!\alpha_{\bf
 k}\Bigl(b_{\bf k\sigma}b_{\bf -k,-\sigma}-b^\dagger_{\bf k\sigma}
 b^\dagger_{\bf -k,-\sigma}\Bigr)\biggr]
\label{ut}\end{equation}
with $\alpha_{\bf k}=\ln[(1+\eta\gamma_{\bf k})/(1-\eta\gamma_{\bf
k})]/4$, $\eta=2/(2-\nu)$, $\gamma_{\bf k}=\sum_{\bf a}\exp(i{\bf
ka})/4$, and the primed sum sign indicates that the summation is
restricted to the magnetic Brillouin zone which is half as large as
the usual one.  In Eq.~(\ref{ut}), $b_{\bf k\sigma}=\sqrt{2/N}\sum_{\bf
m}\exp(-i{\bf km})b_{\bf m}P^\sigma_{\bf m}$ where due to the projector
$P^\sigma_{\bf m}$ the summation is performed over one sublattice.  As a
result, we obtain
\begin{eqnarray}
&&{\cal H}_H=U^\dagger H_HU=\sum_{\bf k\sigma}\!
 \stackrel{\vphantom{a}}{\vphantom{a}}^\prime\!\!\omega^0_{\bf k}
 b^\dagger_{\bf k\sigma}b_{\bf k\sigma},\quad
 \omega^0_{\bf k}=-\frac{4J}{\eta}\langle b_{\bf 0}b_{\bf
 a}\rangle\sqrt{1-\eta^2\gamma^2_{\bf k}},\nonumber\\[-1.5ex]
&&\label{th}\\[-1.5ex]
&&\langle b_{\bf 0}b_{\bf a}\rangle=\frac{2}{N}
 \sum_{\bf k\sigma}\!\stackrel{\vphantom{a}}{\vphantom{a}}^\prime\!\!
 \frac{\gamma_{\bf k}}{\sqrt{1-\eta^2\gamma^2_{\bf k}}}
 \biggl[
 \langle b_{\bf -k,-\sigma}b_{\bf k\sigma}\rangle_U
 -\eta\gamma_{\bf k}\biggl(\langle b^\dagger_{\bf k\sigma}b_{\bf
 k\sigma}\rangle_U+\frac{1}{2}\biggr)\biggr],\nonumber
\end{eqnarray}
where we omitted unessential constant terms, and in $\langle b_{\bf
0}b_{\bf a}\rangle$ for the following discussion we keep the anomalous
correlation $\langle b_{\bf -k,-\sigma}b_{\bf k\sigma}\rangle_U$ which
is nonzero at $x\neq 0$.  The subscript $U$ means that the averaging is
performed with the Hamiltonian transformed with operator (\ref{ut}).  In
the absence of holes we have $\langle b^\dagger_{\bf k\sigma}b_{\bf
k\sigma}\rangle_U=[\exp(\omega^0_{\bf k}/T)-1]^{-1}$ and $\langle
b_{\bf -k,-\sigma}b_{\bf k\sigma}\rangle_U=0$ where $T$ is the
temperature in energy units, while for $x>0$ the correlations are
calculated from the magnon Green's function.  Condition (\ref{zsm}a)
which determines $\eta$ in Eq.~(\ref{th}) acquires the form
\begin{equation}
\frac{2}{N}\sum_{\bf k}\!
 \stackrel{\vphantom{a}}{\vphantom{a}}^\prime\!\!
 \frac{1}{\sqrt{1-\eta^2\gamma^2_{\bf k}}}\biggl(
 \langle b^\dagger_{\bf k\sigma}b_{\bf k\sigma}\rangle_U
 +\frac{1}{2}-\eta\gamma_{\bf k}
 \langle b_{\bf -k,-\sigma}b_{\bf k\sigma}\rangle_U\biggr)=1.
\label{cond}\end{equation}
Analogous equations (without $\langle b_{\bf -k,-\sigma}b_{\bf
k\sigma}\rangle_U$) for the magnon spectrum of the Heisenberg
antiferromagnet without holes were obtained in a somewhat different
manner in Refs.~\cite{takahashi,tang}.  As shown in these works,
in states without long-range antiferromagnetic ordering and for finite
lattices one obtains $\eta<1$ which introduces a gap in the magnon
spectrum (\ref{th}) near the points $(0,0)$ and $(\pi,\pi)$ of the
Brillouin zone.

The reference state of the spin-wave approximation discussed above is
the classical N\'eel state $|{\cal N}\rangle$.  Other states are
described via the reference state and magnon creation operators
determined for this state.  The introduction of holes in this picture
leads to two possibilities for the hole movement:  there is a magnon
or there is no magnon on a site which a hole jumps to.  Both these
possibilities are described by the following term of the Hamiltonian:
\begin{equation}
H_t=t\sum_{\bf ma}h_{\bf m}h^\dagger_{\bf
m+a}\bigl(b_{\bf m+a}+b^\dagger_{\bf m}\bigr),
\label{ht}\end{equation}
which corresponds to the first term in Eq.~(\ref{tjh}).  In
Eq.~(\ref{ht}), $h^\dagger_{\bf m}=\sum_\sigma P^\sigma_{\bf m}|{\bf
m}\rangle \langle{\bf m}\sigma|$ is the hole creation operator in the
N\'eel state $|{\cal N}\rangle=\prod_{\bf m}(\sum_\sigma P^\sigma_{\bf
m}|{\bf m}\sigma\rangle)$.

In spite of the clear physical meaning of the constraint of zero
staggered magnetization, let us discuss the approximations made from a
somewhat different point of view.  Notice that in the case of
short-range antiferromagnetic order characterized by the spin
correlation length $\xi\gg a$ one can use the usual spin-wave
approximation in any crystal region with a linear size of the order of
$\xi$.  A N\'eel state with some local spin quantization axis is used as
the reference state for this spin-wave approximation.  Additionally one
should take into account the finite-size effect due to a finite value
of $\xi$.  This is done by applying condition (\ref{cond}).  The value
of $\eta$ determined by this condition is directly connected with $\xi$
\cite{takahashi} (recall that $\eta<1$ both for a finite $\xi$ and a
finite lattice).  Notice also that only sites neighboring to the hole
site are involved in the processes described by Eq.~(\ref{ht}).  The
hole-magnon interaction is of the short-range type.  This is the reason
why in the case of short-range antiferromagnetic order with $\xi\gg a$
the hole hopping term (\ref{ht}) looks exactly like the same term for
the long-range order \cite{schmitt,kane}.  There is no
``bare-hopping'' (without magnon operators) term in Eq.~(\ref{ht})
because in the case $\xi\gg a$ the direction of the spin quantization
axis is practically the same on sites involved in a hole jump.

After unitary transformation (\ref{ut}) the total Hamiltonian reads
\begin{equation}
{\cal H}=U^\dagger H_tU+{\cal H}_H=\sum_{\bf kk'\sigma}\!\!
 \stackrel{\vphantom{a}}{\vphantom{a}}^\prime\!\!\bigl(g_{\bf kk'}
 h^\dagger_{\bf k\sigma}h_{\bf k-k',-\sigma}b_{\bf k'\sigma}
 +{\rm H.c.}\bigr)
 +\sum_{\bf k\sigma}\!
 \stackrel{\vphantom{a}}{\vphantom{a}}^\prime\!\!\omega^0_{\bf k}
 b^\dagger_{\bf k\sigma}b_{\bf k\sigma},
\label{h}\end{equation}
where $g_{\bf kk'}=-4t\sqrt{2/N}(\gamma_{\bf k-k'}u_{\bf
k'}+\gamma_{\bf k}v_{\bf k'})$ and $u_{\bf k}=\cosh(\alpha_{\bf k})$,
$v_{\bf k}= -\sinh(\alpha_{\bf k})$.  On carrying out the unitary
transformation we neglected the noncommutativity of hole and magnon
operators which is justified at least for $x\lesssim 0.1$ by
comparison with exact diagonalization results \cite{mars}. At
$\eta=1$ Eq.~(\ref{h}) reduces to the Hamiltonian obtained in
Refs.~\cite{schmitt,kane} for the case of the long-range
antiferromagnetic order.  The number of magnons is not conserved by
Hamiltonian (\ref{h}) and at $x>0$ the anomalous correlation $\langle
b_{\bf -k,-\sigma}b_{\bf k\sigma}\rangle_U$ in Eqs.~(\ref{th}) and
(\ref{cond}) is nonzero.

Notice that the used spin-wave approximation is not rotationally
invariant and the correlations $\langle S^+_{\bf l}S^-_{\bf m}\rangle$
are zero \cite{takahashi,tang}.  Therefore only the $z$ components of
spin operators are considered in the following discussion.

\section{Susceptibility and self-energy equations}
In the new notations the spin components (\ref{zso}) acquire the form
\begin{equation}
s^z_{\bf m}=\frac{1}{2}{\rm e}^{i\bf \Pi m}\left(\frac{1}{2}-
 b^\dagger_{\bf m}b_{\bf m}\right),\quad
s^z_{{\bf m+a}/2}=\frac{1}{32}{\rm e}^{i\bf \Pi m}\left[
 h_{\bf m}h^\dagger_{\bf m+a}\left(b_{\bf m+a}-b^\dagger_{\bf m}
 \right)+{\rm H.c.}\right].
\label{zs}\end{equation}
$s^z_{\bf m}$ and $s^z_{{\bf m+a}/2}$ give contributions to the
magnetization
$$M^z_{\bf q}=-2\mu_Bs^z_{\bf q},\quad
s^z_{\bf q}=\sum_{\bf m}s^z_{\bf m}{\rm e}^{-i\bf qm}+
 \sum_{\bf m}\sum_{\bf a'}s^z_{{\bf m+a'}/2}
 {\rm e}^{-i{\bf q(m+a'}/2)},$$
from magnons and holes, respectively.  Here ${\bf a'}=(a,0)$, $(0,a)$.
The susceptibility is determined by the equation
$$\chi^z({\bf q}\omega)=\frac{i}{N}\int^\infty_0d\tau{\rm
 e}^{i\omega\tau}
 \langle\left[M^z_{\bf q\tau},M^z_{\bf -q}\right]\rangle,$$
where $M^z_{\bf q\tau}=\exp(iH\tau)M^z_{\bf q}\exp(-iH\tau)$,
$H=H_t+H_H-\mu{\cal N}$, ${\cal N}=\sum_{\bf k\sigma}\!\!\!\!\!\!
\vphantom{\int}^\prime \,\,\,\, h^\dagger_{\bf k\sigma} h_{\bf
k\sigma}$, and $\mu$ is the hole chemical potential.  The
susceptibility can be calculated using the simplest decoupling, as the
subsequent terms of the perturbation series for the respective
Matsubara Green's function are proportional to powers of the small
hole concentration and are further decreased by rapidly oscillating
coefficients.  In this approximation the magnon and hole contributions
(indicated by subscripts $m$ and $h$, respectively) in the
susceptibility can be written as
\begin{eqnarray}
&&{\rm Im}\,\chi^z_m({\bf q}\omega)=\frac{4\mu^2_B}{N}\sum_{\bf k}
 \int^\infty_{-\infty}\frac{d\nu}{\pi}[n_B(\nu)-
 n_B(\nu+\omega)]\nonumber\\[0.4ex]
&&\quad\quad\times [K_1({\bf k},\nu)K_1({\bf k}+
 \bbox{\kappa},\nu+\omega)+K_2({\bf k},\nu)
 K_2({\bf k}+\bbox{\kappa},\nu+\omega)],\nonumber\\[0.5ex]
&&{\rm Im}\,\chi^z_h({\bf q}\omega)=\frac{\mu^2_B}{16N^2}\sum_{\bf kk'}
 \int\!\!\!\!\int^\infty_{-\infty}\frac{d\nu d\nu'}{\pi^2}
 [n_F(\nu)-n_F(\nu+\nu'-\omega)][n_B(\nu')-n_B(\nu'-\omega)]
 \nonumber\\[0.7ex]
&&\quad\quad\times\:{\rm Im}\,G({\bf k}\nu)\,{\rm Im}\,
 G({\bf k+k'-q},\nu+\nu'-\omega) \label{chi}\\[0.6ex]
&&\quad\quad\times\left[{\rm Im}\,D_{11}({\bf k'}\nu')\,\gamma^2\!\!
 \left({\bf k}-\frac{\bf q}{2}\right)-{\rm Im}\,D_{11}({\bf k'},-\nu')\,
 \gamma^2\!\!\left({\bf k+k'}-\frac{\bf q}{2}\right)\right.
 \nonumber\\[0.4ex]
&&\quad\quad\left. -2{\rm Im}\,D_{12}({
 \bf k'}\nu')\,\gamma\!\left({\bf k}-\frac{\bf q}{2}\right)\gamma\!
 \left({\bf k+k'}-\frac{\bf q}{2}\right)\right],\nonumber\\[0.8ex]
&&{\rm Re}\,\chi({\bf q}\omega)={\cal P}\int^\infty_{-\infty}
 \frac{d\nu}{\pi}\frac{{\rm Im}\,\chi({\bf q}\omega)}{\nu-\omega},
 \nonumber
\end{eqnarray}
where $n_B=[\exp(\omega/T)-1]^{-1}$, $n_F=[\exp(\omega/T)+1]^{-1}$,
$\bbox{\kappa}={\bf q}-{\bf\Pi}$,
\begin{eqnarray*}
K_1({\bf k}\omega)&=&u^2_{\bf k}{\rm Im}\,D_{11}({\bf k}\omega)+
 2u_{\bf k}v_{\bf k}{\rm Im}\,D_{12}({\bf k}\omega)
 -v^2_{\bf k}{\rm Im}\,D_{11}({\bf k},-\omega),\\[0.75ex]
K_2({\bf k}\omega)&=&u_{\bf k}v_{\bf k}[{\rm Im}\,D_{11}({\bf k}\omega)-
 {\rm Im}\,D_{11}({\bf k},-\omega)]
 +(u^2_{\bf k}+v^2_{\bf k}){\rm Im}\,D_{12}({\bf k}\omega),
\end{eqnarray*}
and $D_{ij}({\bf k}\omega)$, $G({\bf k}\omega)$ are the Fourier
transforms of the magnon and hole Green's functions
\begin{eqnarray*}
D_{11}({\bf k}t)=-i\theta(t)\langle[&&b_{\bf k\sigma}(t),
b^\dagger_{\bf k\sigma}]\rangle_U,\quad
D_{12}({\bf k}t)=-i\theta(t)
\langle[b_{\bf -k,-\sigma}(t),b_{\bf k\sigma}]\rangle_U,\\[0.75ex]
&&G({\bf k}t)=-i\theta(t)\langle[h_{\bf k\sigma}(t),h^\dagger_{\bf
k\sigma}]\rangle_U,
\end{eqnarray*}
$b_{\bf k\sigma}(t)=\exp[i({\cal H}-\mu{\cal N})t]b_{\bf k\sigma}
\exp[-i({\cal H}-\mu{\cal N})t]$.  These Green's functions do not
depend on the spin index.  In Eq.~(\ref{chi}), summations over wave
vectors are performed over the full Brillouin zone.  In the second
magnetic zone, which together with the first magnetic zone forms this
full zone, $D_{11}({\bf k}\omega)$ and $G({\bf k}\omega)$ repeat
periodically their values in the first zone, while $D_{12}({\bf
k}\omega)$ changes sign.

These Green's functions are determined from the set of self-energy
equations.  The hole $\Sigma$ and magnon $\Pi_{11}$, $\Pi_{12}$
self-energies are described by the following diagrams:
\centerline{\psfig{figure=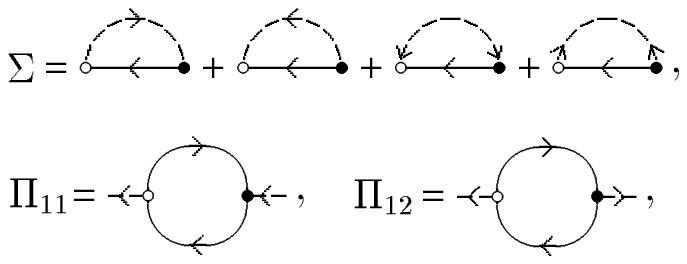,height=3.5cm}}
where solid and dashed lines correspond to hole and magnon Green's
functions, open and filled circles are bare and full vertices.  Dashed
lines with two oppositely directed arrows correspond to the anomalous
magnon Green's functions $D_{12}({\bf k}t)$ and $D_{21}({\bf
k}t)=-i\theta(t) \langle[b^\dagger_{\bf k\sigma}(t),b^\dagger_{\bf
-k,-\sigma}]\rangle_U$.  The first correction to the bare vertex
$g_{\bf kk'}$ is exactly zero due to the impossibility to coordinate
spin indices in this correction \cite{krier}.  This suggests the use
of the Born approximation in which the full vertices are substituted
with the bare ones.  In this approximation the real-frequency
self-energy equations read
\begin{eqnarray}
&&G({\bf k}\omega )=\bigl[\omega +\mu -
    \Sigma ({\bf k}\omega)\bigr]^{-1}\! ,\nonumber\\[0.4ex]
&&{\rm Im}\,\Sigma ({\bf k}\omega )=-2\sum_{\bf k'}\!
    \stackrel{\vphantom{a}}{\vphantom{a}}^\prime\!\!
    \int^\infty_{-\infty}\frac{d\omega '}{\pi}\Bigl[
    g^2_{\bf kk'}{\rm Im}\, D_{11}({\bf k'}\omega ')
    -g^2_{\bf k-k',-k'}{\rm Im}\, D_{11}(-{\bf k'},-\omega ')
    \nonumber\\
&&\quad\quad\quad +2g_{\bf kk'}g_{\bf k-k',-k'}{\rm Im}\,
    D_{12}({\bf k'}\omega ')
    \Bigr]\bigl[ n_B(\omega ')+n_F(\omega '-\omega )\bigr]
    {\rm Im}\, G({\bf k}-{\bf k'},\omega -\omega '),
    \nonumber\\[0.8ex]
&&{\rm Re}\,\Sigma ({\bf k}\omega )={\cal P}
    \int^\infty_{-\infty}\frac{d\omega '}{\pi}\,\frac{{\rm Im}\,
    \Sigma ({\bf k}\omega ')}{\omega '-\omega},\nonumber\\[-1.2ex]
&&\label{se}\\[-1.2ex]
&&D_{11}({\bf k}\omega )=\frac{R^*({\bf k},-\omega)}
    {R({\bf k}\omega)R^*({\bf k},-\omega)-
    \Pi^2_{12}({\bf k},\omega)},\quad
    D_{12}({\bf k}\omega )=\frac{\Pi_{12}({\bf k},\omega)}
    {R({\bf k}\omega)R^*({\bf k},-\omega)-
    \Pi^2_{12}({\bf k},\omega)},\nonumber\\[0.5ex]
&&{\rm Im}\,\Pi_{11}({\bf k}\omega )=
    2\sum_{\bf k'}\!
    \stackrel{\vphantom{a}}{\vphantom{a}}^\prime\!\!
    g^2_{\bf k'k}\int^\infty_{-\infty}
    \frac{d\omega '}{\pi}\, {\rm Im}\, G({\bf k'}\omega ')\,
    {\rm Im}\, G({\bf k'}-{\bf k},\omega '-\omega )
    \bigl[ n_F(\omega ')-n_F(\omega '-\omega )\bigr] ,
    \nonumber\\
&&{\rm Im}\,\Pi_{12}({\bf k}\omega )=
    2\sum_{\bf k'}\!
    \stackrel{\vphantom{a}}{\vphantom{a}}^\prime\!\!
    g_{\bf k'k}g_{\bf k'-k,-k}\int^\infty_{-\infty}
    \frac{d\omega '}{\pi}\, {\rm Im}\, G({\bf k'}\omega ')\,
    {\rm Im}\, G({\bf k'}-{\bf k},\omega '-\omega )\nonumber\\
&&\quad\quad\quad\times\bigl[ n_F(\omega ')-n_F(\omega '-\omega )
    \bigr] ,\nonumber\\[0.4ex]
&&{\rm Re}\,\Pi_{ij}({\bf k}\omega )={\cal P}
    \int^\infty_{-\infty}\frac{d\omega '}{\pi}\,
    \frac{{\rm Im}\,\Pi_{ij}({\bf k}\omega ')}{\omega '-\omega},
    \nonumber
\end{eqnarray}
where $R({\bf k}\omega )=\omega -\omega^0_{\bf k} -\Pi_{11}({\bf
k}\omega)$.  Equations (\ref{se}) and (\ref{cond}) form a
self-consistent set and can be solved iteratively.  The calculation
procedure is the following:  for given values\hfill of\hfill
$\mu$\hfill and\hfill $T$\hfill some\hfill starting\hfill value\hfill
$\eta<1$\hfill was\hfill selected;\hfill after\hfill the\hfill
convergence\hfill of

\vspace{3mm}
\hbox{
\psfig{figure=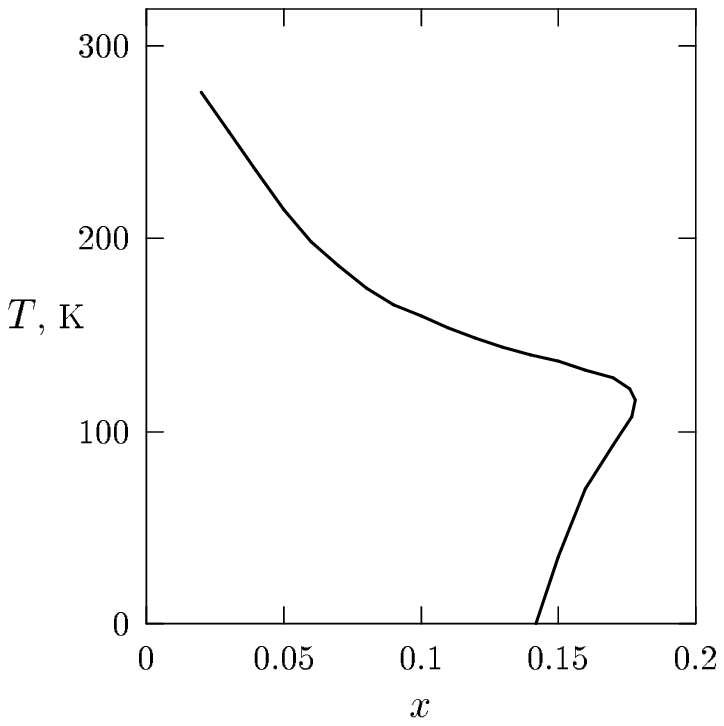}\hspace*{2mm}
\raisebox{4cm}{\parbox{7.4cm}{\small Fig.~1.~The curve encloses the
region where condition (\ref{cond}) can be fulfilled.}}}

\noindent the iterations condition (\ref{cond}) is checked and $\eta$
is appropriately changed for the next iteration cycle, until the
condition is fulfilled with the accuracy of 10$^{-3}$ (some other
details of the calculation procedure can be found in
Ref.~\cite{sher97a}).  In the calculations a 20$\times$20 lattice was
used.  Green's functions were computed on a mesh of frequency points
equally spaced with the step $\Delta\omega\approx 0.022t\approx
11$~meV.  We found that condition (\ref{cond}) can be satisfied only
in a limited region of the $T$-$x$ plane, namely below the curve shown
in Fig.~1.

\section{The energy spectrum}
The evolution of the calculated hole spectral function $A({\bf
k}\omega)=-{\rm Im}\,G({\bf k}\omega)$ with the concentration
\begin{equation}
x=-\frac{2}{\pi N}\sum_{\bf k}\int_{-\infty}^{\infty}d\omega\,
n_F(\omega )\,{\rm Im}\,G({\bf k}\omega)
\label{conc}\end{equation}
is shown in Fig.~2 for two points of the Brillouin zone.  We use the
hole picture where

\centerline{\psfig{figure=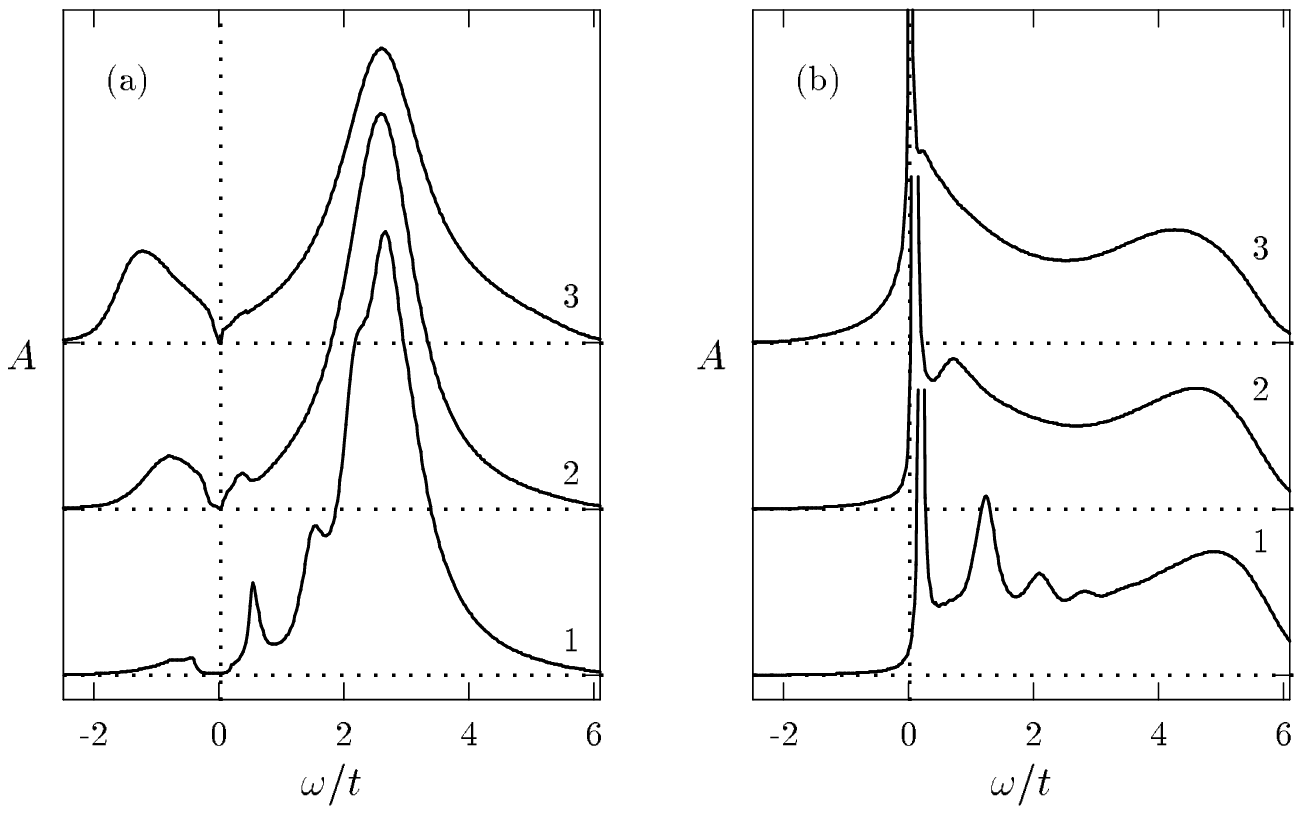,height=6.5cm}}

\vspace*{14mm}\noindent{\small Fig.~2.~The hole spectral function
$A({\bf k}\omega)$ for ${\bf k}=(0,0)$ (a) and $(0,\pi)$ (b).  $T=0$.
Curves 1, 2, and 3 correspond to $x=0.016$, $0.059$, and $0.133$,
respectively.}

\vspace{3ex}
\noindent states below the Fermi level
$\omega=0$ are filled by holes.  Since our description is based on the
spin-wave approximation with the magnetic Brillouin zone which is twice
smaller than the full Brillouin zone, the spectral functions in the
points ${\bf k}$ and ${\bf k}+(\pi,\pi)$ are identical.  For long-range
antiferromagnetic ordering when these points are equivalent such
description does not lead to any loss of information.  In the case of
short-range order the use of the smaller Brillouin zone leads to a
somewhat coarsened description.  The points ${\bf k}$ and ${\bf
k}+(\pi,\pi)$ are no longer equivalent --- the respective spectral
functions contain similar maxima which however have essentially
different intensities and widths \cite{kampf}.  In our approach, where
the maxima in the two points appear together in one spectral function
for the momentum in the first magnetic Brillouin zone, we cannot
determine whether a selected maximum is more intensive in the first or
in the second magnetic Brillouin zone (which form the full zone).  In
other words we cannot determine where is the usual band and where is the
shadow band \cite{kampf}.  However, we can distribute the maxima between
the two magnetic zones by drawing additional information from experiment
as we shall do in the following discussion.

At $x\lesssim 0.04$ the spectra contain series of peaks and the
hole-magnon scattering continuum (see curves 1 in Fig.~2).  With
increasing $x$ only the lowest and most intensive of these peaks is
retained in the spectrum for wave vectors near the boundary of the
magnetic Brillouin zone [everywhere in this region the spectrum is
similar to that for ${\bf k}=(0,\pi)$].  This peak corresponds to the
so-called spin-polaron band in the hole spectrum.  Other peaks are
washed away, forming a broad dispersive maximum above the Fermi level
for momenta in the central part of the magnetic Brillouin zone.
Simultaneously a maximum develops below the Fermi level (compare
curves 2 and 3 in Fig.~2).  As seen in Fig.~3, this maximum possesses
dispersion and is comparatively narrow near the Fermi level.

The energy positions of the spectral maxima discussed above as
functions of momentum are shown in Fig.~4 for heavily underdoped and
moderately doped cases \hfil [as

\vspace{2mm}
\hbox{
\psfig{figure=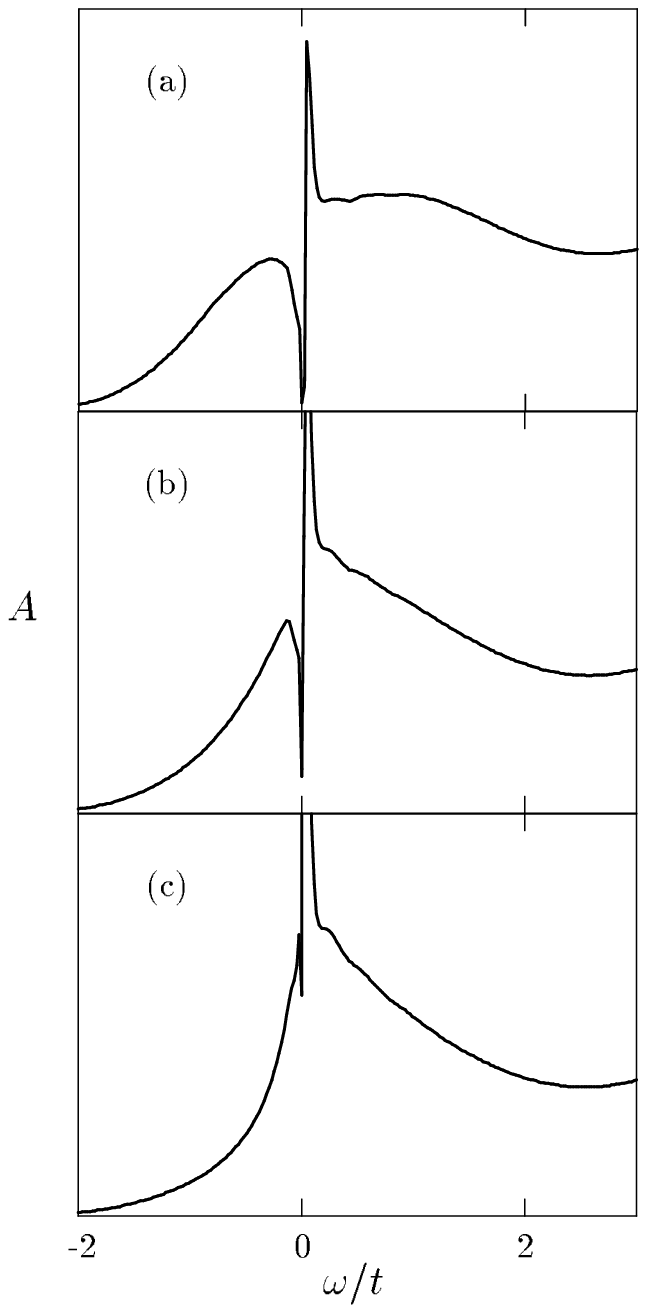}
\hspace{-2mm}
\raisebox{7cm}{\parbox{7.8cm}{\small Fig.~3.~The hole spectral
function for ${\bf k}=(0.5\pi,\pi)$ (a), $(0.4\pi,\pi)$ (b), and
$(0.3\pi,\pi)$ (c).  $x=0.133$, $T=0$.}}}

\hspace*{2mm}\psfig{figure=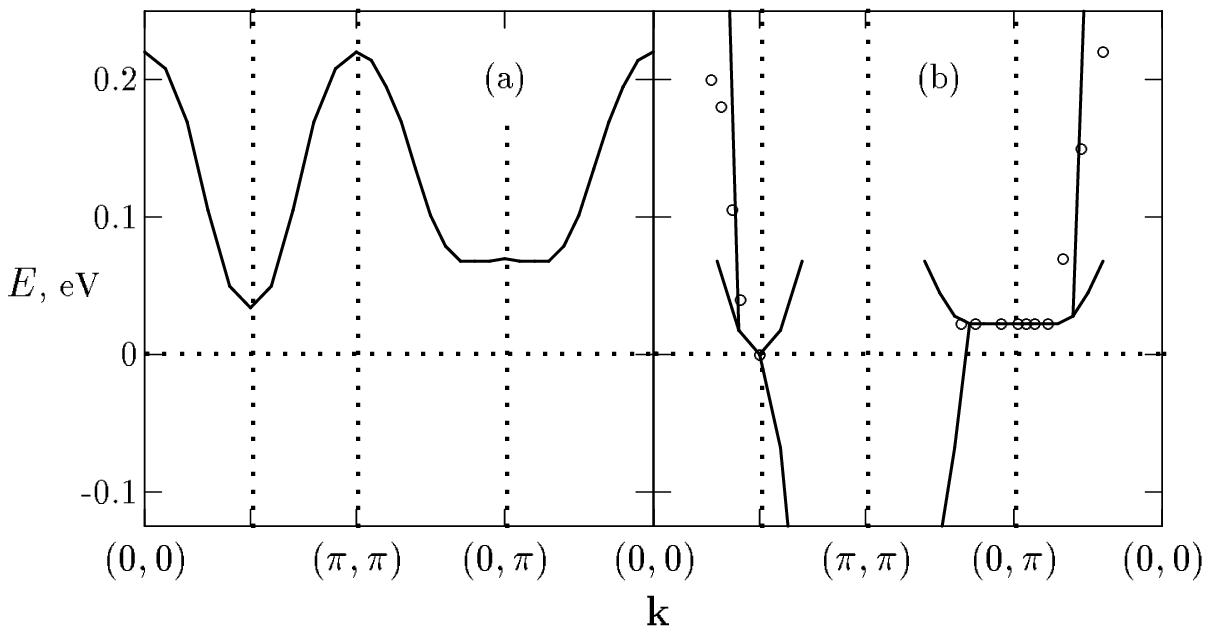}

\noindent{\small Fig.~4.~Energy vs.\ momentum relationships calculated
for $x=0.021$, $T=0$ (a) and $x=0.121$, $T=0.02t=116$~K (b, solid
line).  In part (b), circles indicate positions of quasiparticle peaks
in the photoemission experiment \protect\cite{shen} carried out at
$T=100$~K in a Bi2212 crystal with $T_c=85$~K.}

\vspace{3ex}
\noindent discussed above, we have used the experimental fact that
hole pockets are positioned around $(\pm\pi,\pm\pi)$ to relate the
portion of the energy band below the Fermi level to this part of the
full Brillouin zone].  Figure~4a demonstrates the well-known
spin-polaron band discussed in a large number of works devoted to the
case of low doping (see, e.g.,
\cite{mars,krier,sher90,martinez,plakida,sher94}).  The width of this
band is of the order of $J$, i.e.\ much less than the hole hopping
constant $t$.  This is connected with the above discussed fact that in
the rigorous antiferromagnetic order the hole movement requires the
emission and absorption of magnons.  In such conditions the slower
subsystem --- magnons --- will determine the bandwidth of the combined
quasiparticle.  As mentioned, a part of the spin-polaron band is
retained near the boundary of the magnetic Brillouin zone for moderate
doping.  As seen in Fig.~4b, this part retains, with some distortion,
main features of the spin-polaron band in this region.  Among these
features is the large nearly flat region which is visible around
$(0,\pi)$ in Fig.~4b.  In this figure we have compared the calculated
energy band with the normal-state photoemission data \cite{shen} in
Bi2212 with $T_c=85$~K which corresponds to the hole concentration in
the range $0.11-0.15$.  In Ref.~\cite{shen} the energy position of the
flat region --- the so-called extended van Hove singularity --- is
estimated to be $\pm 30-50$~meV relative to the Fermi level.  We
positioned the experimental flat region at $+20$~meV (in the hole
picture), in accordance with later more exact measurements
\cite{ding}.  As seen in Fig.~4b, with respect to both the energy
position and the extension in the Brillouin zone the calculations
reproduce well the experimental van Hove singularity.  Notice that
also the general shape of the band, as seen in photoemission (which
tests the region $\omega\geq 0$), is well reproduced by the
calculations.  That the flat portion of the spin-polaron band can
correspond to the photoemission extended saddle point was apparently
first indicated in Ref.~\cite{dagotto}.

As follows from Fig.~4, considerable changes occur in the hole
band shape on moving from light to moderate dopings, as was first
indicated in Refs.~\cite{sher94,sher97a}.  The narrow, with the width of
the order of $J$, spin-polaron band is transformed to a much wider band
characterized by the energy parameter $t$.  The general shape of this
band resembles a considerably distorted 2D nearest-neighbor band
produced by the kinetic term of the $t$-$J$ Hamiltonian (\ref{tjh}).
Analogous changes in the band shape are observed in the photoemission of
Bi2212 \cite{marshall}.  These changes point to a certain weakening of
correlations for $x>0.04$, however, some features of the strongly
correlated spectrum are retained:  as mentioned, near the Fermi level
around the boundary of the magnetic Brillouin zone the spectrum contains
a persistent part of the spin-polaron band and widths of spectral maxima
grow steeply with distance from the Fermi level.

These changes in the hole spectrum are connected with changes in the
magnon subsystem.  For $T=0$ starting from $x\approx 0.02$ in the
central part of the magnetic Brillouin zone magnons become
overdamped.  The overdamped magnons manifest themselves in a
perceptible intensity of the magnon spectral function $B_{11}({\bf
k}\omega)=-{\rm Im}\,D_{11}({\bf k}\omega)$ in the nearest vicinity
and on both sides of $\omega=0$ (see Fig.~5a where together with the
structure corresponding to an overdamped magnon the maximum of a usual
magnon is also visible; notice that overdamped magnons do not form
maxima in $B_{11}$ because this quantity changes sign at the central
frequency $\omega=0$ of these magnons).  The appearance of the
overdamped magnons points to the destruction of the long-range
antiferromagnetic order by holes \cite{sher93a,onufrieva} which, in
contrast to the destruction due to thermal fluctuations, occurs also
at zero temperature.  Due to a finite intensity in $B_{11}$, produced
by these magnons in the range $\omega<0$, the magnon occupation
number $n_{\bf k}=-\int^\infty_{-\infty}d\omega \pi^{-1}n_B(\omega)
B_{11}({\bf k}\omega)$ is finite at $T=0$ which leads to a finite
zero-temperature correlation length $\xi$ in the spin correlation
function $\langle s^z_{\bf l}s^z_{\bf m}\rangle$ (in the considered
finite lattice this phase transition is smeared; fortunately for
relevant small $x$ an analytic consideration for an infinite lattice
is possible \cite{sher93a,onufrieva}).  The overdamped magnons can be
identified with relaxational modes describing relative rotations of
magnetic quantization axes in regions of size $\xi$.  In the hole
spectrum the destruction of the long-range antiferromagnetic order
manifests itself in the mentioned change of the shape and
characteristic energy of the spectrum from $J$ to $t$ at $x\approx
0.04$.  Qualitatively this \hfil can \hfil be \hfil understood \hfil
in \hfil the \hfil following \hfil way:  \hfil after \hfil the \hfil
destruction \hfil of \hfil the \hfil long-range

\vspace*{1.5ex}\hspace*{-1mm}\psfig{figure=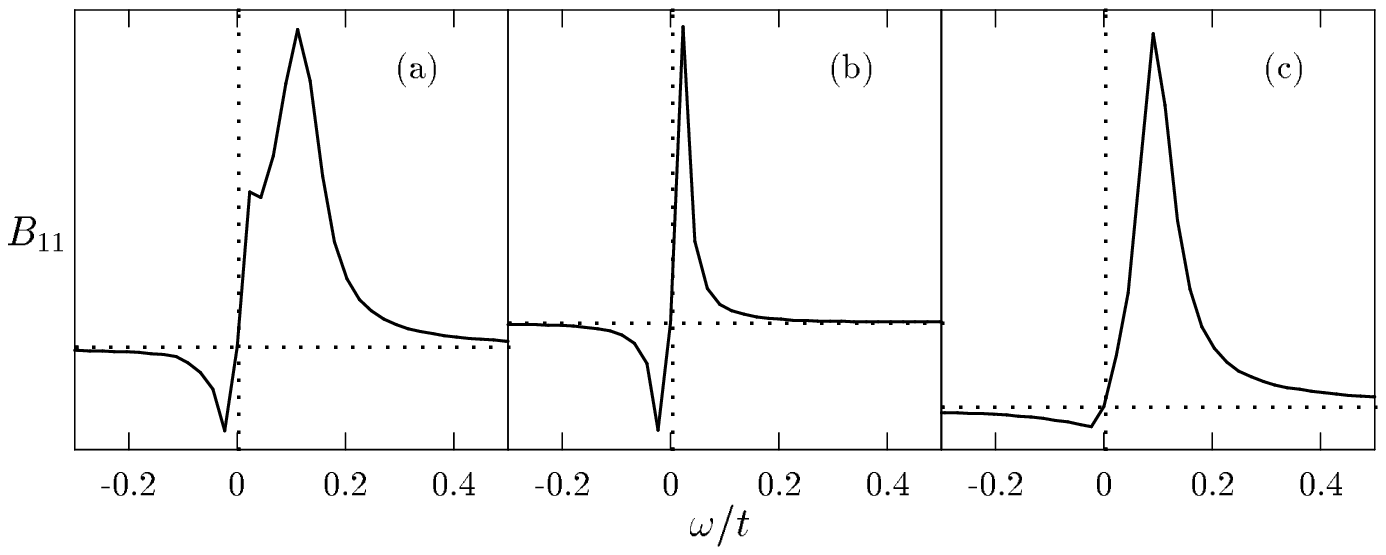}

\noindent{\small Fig.~5.~The magnon spectral function $B_{11}({\bf
k}\omega)$ for ${\bf k}=(0,\pi/5)$, $T=0$, $x=0.059$ (a), ${\bf
k}=(0,0)$, $T=58$~K, $x=0.1$ (b), and ${\bf k}=(\pi/2,\pi/2)$,
$T=116$~K, $x=0.172$ (c).}

\noindent order holes can move without introducing additional disorder
in the magnon subsystem and, as a result, the larger characteristic
energy, $t$, reveals itself in the spectrum.

Apparently nearly simultaneously with the appearance of overdamped
magnons $\eta$ becomes finite and a gap opens in the magnon spectrum.
The low-frequency overdamped magnons transform this gap to a pseudogap.
This pseudogap is most evident in $B_{11}({\bf k}=0,\omega)$ by the
asymmetry of the spectrum around $\omega=0$ (Fig.~5b).  In spite of the
appearance of overdamped magnons, even at $x\approx 0.17$ and $T\approx
100$~K usual magnons with essentially softened frequencies and increased
damping are retained at the periphery of the magnetic Brillouin zone
indicating persistent antiferromagnetic correlations (see Fig.~5c).

Let us return to the hole spectrum.  The hole Fermi surfaces
calculated for moderate dopings are shown in Figs.~6a and 6b.  In
part~(a), line segments along the boundary of the magnetic Brillouin
zone are connected with the spin-polaron band.  After touching the
Fermi level (which occurs at $x\approx 0.01$) the bottom of this band
remains pinned to the Fermi level and the band flattens with
increasing $x$.  Originating from work \cite{shraiman} it is widely
believed that for small hole concentrations the Fermi surface consists
of small hole pockets around $(\pm\pi/2,\pm\pi/2)$.  Our calculations
do not support this point of view.  The Fermi level does not cross the
bottom of the spin-polaron band but is rather pinned to this bottom
and the mentioned hole pockets do not arise [notice however that the
line segments near $(\pm\pi/2,\pm\pi/2)$ in Fig.~6 can have a finite
width provided that it is much less than our momentum resolution
$\pi/10$].

\vspace{2ex}
\hspace*{-1mm}\psfig{figure=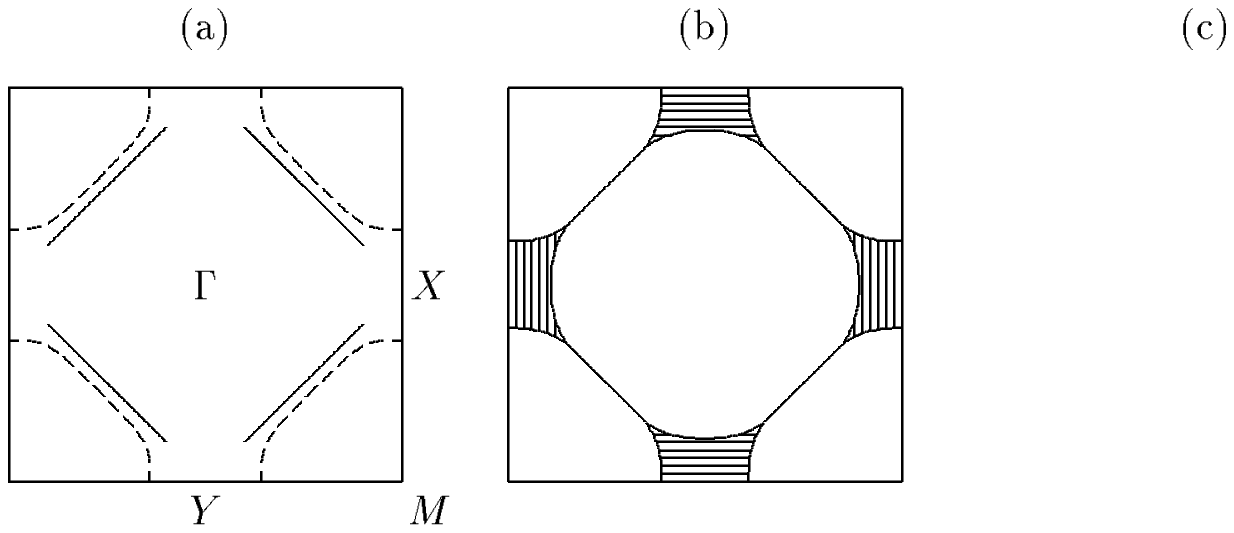}

%\vspace{4cm}
\noindent{\small Fig.~6.~The calculated hole Fermi surface for
$T\approx 100$~K, $0.07\lesssim x < 0.17$ (a), $x=0.172$ (b), and
experimental Fermi surface for several cuprates (c, from
Ref.~\cite{king}).  Dashed lines in (a) and hatched regions in (b)
indicate hidden and two-dimensional parts of the surface.  The points
$X$, $Y$, and $M$ correspond to ${\bf k}=(\pi,0)$, $(0,-\pi)$, and
$(\pi,\pi)$, respectively.}

\vspace{3ex}
In Fig.~6a, dashed curves indicate the crossings of the Fermi level by
the part of the energy band which arises with growing $x$ below the
bottom of the spin-polaron band (see Figs.~2--4).  This part forms
hole pockets around $(\pm\pi,\pm\pi)$ and determines the hole
concentration (\ref{conc}).  The pockets start to form from small $x$
and their extent in the Brillouin zone grows with $x$; however up to
$x\approx 0.07$ the pockets do not reach the Fermi level --- all their
states are positioned below the Fermi level.  Thus, for $x\lesssim
0.07$ the Fermi surface consists of only the line segments along the
boundary of the magnetic Brillouin zone and no closed Fermi surface
exists.  For larger hole concentrations the hole pockets reach the
Fermi level and the crossings depicted by the dashed curves in Fig.~6a
arise.  However, as seen in Fig.~3, just at this crossing the weaker
maximum corresponding to the pocket is completely lost within the foot
of a much more intensive spin-polaron peak which is located somewhat
above the Fermi level for the temperature, hole concentration and wave
vectors of Fig.~3.  Thus, the dashed curves in Fig.~6a correspond to
hidden parts of the Fermi surface.  If like in photoemission
experiments the spectrum is not tested below the Fermi level (in the
hole picture), the respective crossings reveal themselves only as a
finite spectral intensity at the Fermi level.

As mentioned, with growing $x$ the persistent part of the spin-polaron
band flattens.  As a consequence, the extended saddle points approach
the Fermi level and at $x=0.172$ they fall on it.  The Fermi surface of
the considered two-dimensional fermions becomes two-dimensional, as
shown in Fig.~6b.  The mentioned hole concentration is close to
optimal doping which corresponds to the highest $T_c$.  As observed in
photoemission of Bi2212 \cite{marshall}, in this case the saddle points
do lie, within the experimental accuracy, on the Fermi level.
Fig.~6c demonstrates Fermi surfaces deduced from photoemission in
Nd$_{2-x}$Ce$_x$CuO$_4$ (NCCO), Bi2212, and $\rm
Bi_2(Sr_{0.97}Pr_{0.03})_2CuO_{6+\delta}$ (Bi2201).  Contrasting this
figure with Figs.~6a and 6b, we conclude that our calculations
reproduce satisfactorily the main features of the experimental Fermi
surface in crystals with significantly different hole concentrations.

For Fig.~6a we indicated the concentration range where the Fermi surface
remains practically unchanged.  In particular, it means that the size of
the $(\pm\pi,\pm\pi)$ hole pockets varies only slightly in the
concentration range $0.07\lesssim x\lesssim 0.17$.  This result of our
calculations agrees with experiment \cite{loeser} and points to the
substantial violation of Luttinger's theorem \cite{luttinger}.  In
accord with this theorem the area enclosed by the Fermi surface in the
Brillouin zone should vary linearly with $x$.  As seen in Fig.~2,
quasiparticle weights of states (spectral intensities) which form the
hole pockets grow with increasing $x$.  It is this growth of
quasiparticle weights, rather than the growing size of these pockets,
which leads to the increase of the hole concentration in
Eq.~(\ref{conc}).

As follows from the above discussion, the energy spectrum of the
considered fermions differs essentially from the Fermi liquid behavior
of conventional metals.  To this we can add the linear, rather than
quadratic, frequency dependences of the hole decay widths near the Fermi
level in the case of moderate doping \cite{sher94}.  This result
resembles the marginal Fermi liquid \cite{varma}, however, in contrast
to this concept our calculated ${\rm Im}\,\Sigma(\omega)$ has markedly
different slopes below and above the Fermi level and a strong momentum
dependence.

\section{Pseudogap in the hole spectrum}
As follows from Fig.~3, the spin-polaron peak is the most prominent
feature of the hole spectrum near the Fermi level for wave vectors in
the vicinity of the Fermi surface.  The hole spectral function is
directly related to the photoemission spectrum.  Hence the
spin-polaron peak will determine the position of the leading edge of
the photoemission

\hbox{
\psfig{figure=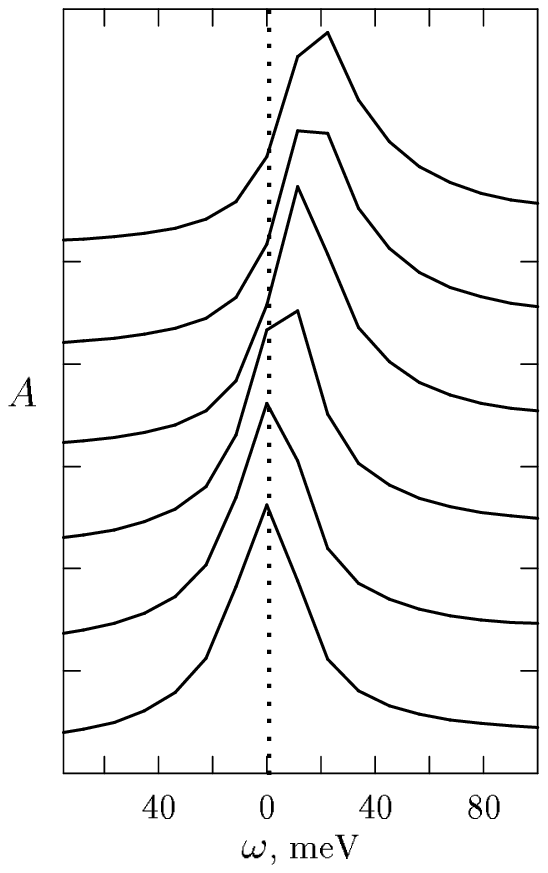}
\hspace{-2mm}
\raisebox{4.5cm}{\parbox{8.4cm}{\small Fig.~7.~The hole spectral
function for wave vectors on the Fermi surface.  Curves from top to
bottom correspond to ${\bf k}=(0.2\pi,\pi)$, $(0.2\pi,0.9\pi)$,
$(0.2\pi,0.8\pi)$, $(0.3\pi,0.7\pi)$, $(0.4\pi,0.6\pi)$, and
$(0.5\pi,$ $0.5\pi)$, respectively.  $T=116$~K, $x=0.121$.}}}

\noindent spectrum.  For several wave vectors on the Fermi surface the
hole spectral function is shown in Fig.~7 for the underdoped case.  As
seen from this figure, the spin-polaron maximum, which lies on the
Fermi level near $(\pi/2,\pi/2)$, is shifted upwards on approaching
$(\pi/5,\pi)$.  Recall that the part of the Fermi surface near
$(\pi/5,\pi)$ is connected with another, broader and weaker maximum
which is completely lost at the foot of the spin-polaron peak on
crossing the Fermi level (see Fig.~3).  As a consequence, the
situation in Fig.~7 looks like a part of the Fermi surface disappears
and a gap opens between the hole energy band and the Fermi level
near $(\pi/5,\pi)$.  Due to the mentioned hidden crossing of the Fermi
level in this point there remains a finite intensity at $\omega=0$
which transforms the gap to the pseudogap.  The same behavior of the
leading edge is observed in photoemission of underdoped cuprates which
led to the idea of the photoemission pseudogap
\cite{loeser,marshall,ding}.

In Fig.~8 we compare our calculated position of the spin-polaron peak
and the experimentally measured position of the photoemission leading
edge \cite{ding} as functions of momentum along the Fermi surface
shown in the insets of both parts of the figure (some differences in
the Fermi surfaces shown may be connected both with the experimental
resolution which produces some uncertainty in the position of the
Fermi surface especially in the case of narrow bands and with the
influence of terms not included in the $t$-$J$ Hamiltonian).  In part
(b), the location on the Fermi surface is determined by the angle
measured from the line $(\pi,\pi)-(\pi,0)$.  At $T=14$~K experimental
curves for the 83~K and 87~K samples in Fig.~8b correspond to
superconducting gaps.  As indicated in Ref.~\cite{ding}, in the
former, underdoped sample the shape of the curve and the magnitude of
the gap remain practically unchanged when the temperature increases
and somewhat exceeds $T_c$, while the gap is closed in the latter,
optimally doped sample.  Taking into account these experimental facts
we conclude from the comparison of Figs.~8a and 8b that the
calculations reproduce satisfactorily the general shape and the
magnitude of

\hspace*{3cm}(a)\hspace{7cm}(b)

\hspace*{0.2cm}\psfig{figure=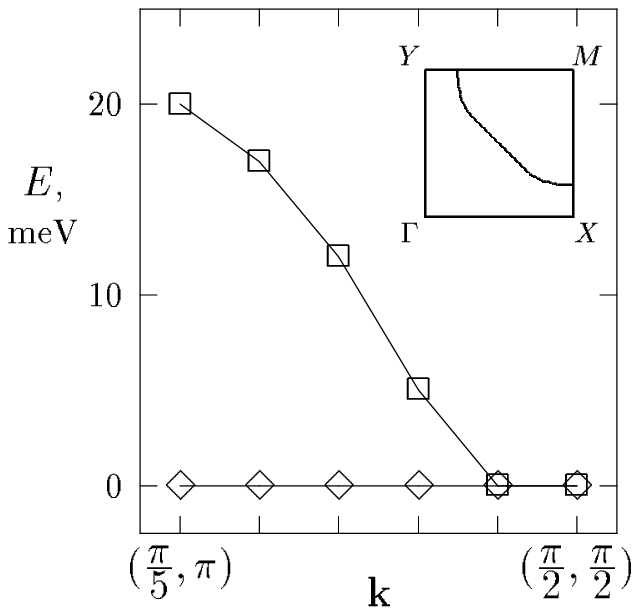,height=6.5cm}

\vspace*{-1ex}
\noindent{\small Fig.~8.~(a)~The position of the spin-polaron peak
along the Fermi surface, shown in the inset, for $x=0.121$ ($\Box$)
and $x=0.172$ ($\Diamond$) at $T=116$~K.  The points $M$, $X$, and $Y$
are defined in the caption of Fig.~6.  (b) The position of the leading
edge of the photoemission spectrum vs.\ momentum along the Fermi
surface in the inset for samples with $T_c=10$~K (heavily underdoped,
triangles), 83~K (underdoped, squares), and 87~K (optimally doped,
circles).  Measurements in Ref.~\cite{ding}, where this figure were
taken from, were carried out for Bi2212 at $T=14$~K\@.  The points
$\overline{M}$ and $Y$ correspond to the $X, Y$ and $M$ points in the
square Brillouin zone.}

\vspace{3ex}
\noindent the normal-state pseudogap in the underdoped case.
Moreover, as seen from Fig.~8a, in agreement with experiment the
pseudogap is closed in the case of optimal doping.

In Ref.~\cite{ding} the similarity of symmetries of the
superconducting gap and the normal-state pseudogap and their smooth
evolution into each other with temperature served as the basis for the
supposition that the pseudogap is the normal-state precursor of the
superconducting gap.  In accord with this supposition the pseudogap
arises above $T_c$ due to superconducting fluctuations.  We did not
include these fluctuations in our present calculations and thus the
pseudogap shown in Fig.~8a is not connected with the fluctuations.  In
our calculations the pseudogap arises due to the specific dispersion
of the spin-polaron band, a part of which is retained near the Fermi
level at moderate doping and gives the most intensive maxima in the
spectral function.  The existence of this band is a consequence of
strong electron correlations.  We do not exclude the possibility that
the superconducting fluctuations do contribute to the pseudogap,
however, based on the satisfactory agreement between our estimate and
the observed pseudogap magnitude we suppose that the main contribution
is provided by strong electron correlations.  Besides, with the
superconducting fluctuation mechanism it is difficult to understand
why the pseudogap appears only on one side of the Fermi level and why
the fluctuations disappear abruptly in a small concentration range
near optimal doping when the pseudogap is closed.  For the 2D $t$-$J$
model various calculations (see, e.g.,
Refs.~\cite{sher97a,scalapino,sher95,plakida97}) give the
$d_{x^2-y^2}$ symmetry for the superconducting gap in the case of
moderate doping.  This gap looks similar to the pseudogap in Fig.~8a,
as the symmetries of both of them are determined by the hole-magnon
interaction and by short-range antiferromagnetic ordering.  This
ensures smooth evolution of the pseudogap into the superconducting gap
with lowering temperature below $T_c$.

As observed in Ref.~\cite{ding}, in slightly underdoped samples the
pseudogap is closed when temperature exceeds some $T^*$.  This
characteristic temperature increases steeply with decreasing hole
concentration.  We believe that this behaviour is connected with
crossing the boundary shown in Fig.~1.  Outside of this boundary for
$x\gtrsim 0.12$ we found a hole energy spectrum which is similar to
the spectrum of the usual metal and does not contain any pseudogap.

Concluding this section let us mention two differences we see between
our calculated and the experimental results.  First, in our calculations the
magnitude of the pseudogap increases slightly with decreasing hole
concentration, while in the experiment \cite{ding} it remains unchanged,
within experimental errors, for samples with markedly different $x$ (see
Fig.~8b).  Second, in the normal-state photoemission spectra of
underdoped Bi2212 the linewidth increases dramatically on moving from
$(\pi/2,\pi/2)$ to the vicinity of $(0,\pi)$ \cite{marshall}.  Though
the photoemission line shape is not well understood \cite{liu}, an
analogous behavior of linewidths could be expected in the hole spectral
function.  However, in our results in Fig.~7 the change of the linewidth
does not look so dramatic.  This contradiction points either to the
experimental pseudogap being somewhat larger than that reported in
Ref.~\cite{ding} [in this case the overdamped magnons wash away the
$(0,\pi)$ maximum, as they did with the $(0,0)$ maximum in Fig.~2a,
curve 2] or to some decay process with low-frequency excitations not
included in the $t$-$J$ model.

\section{Magnetic properties}
The spin correlation function $C_{\bf l-m}=\langle s^z_{\bf l}s^z_{\bf
m}\rangle$ is given by
\begin{mathletters}
\label{sc}
\begin{equation}
C_{\bf l-m}=-\frac{1}{4}\delta_{\bf ml}+
 \biggl\{\frac{2}{N}\sum_{\bf k}\!
 \stackrel{\vphantom{a}}{\vphantom{a}}^\prime\!\!
 \frac{{\rm e}^{i\bf k(m-l)}}{\sqrt{1-\eta^2\gamma^2_{\bf
 k}}}\biggl[\langle
 b^\dagger_{\bf k\sigma}b_{\bf k\sigma}\rangle_U+\frac{1}{2}-\eta
 \gamma_{\bf k}\langle b_{\bf -k,-\sigma}b_{\bf k\sigma}\rangle_U
 \biggr]\biggr\}^2
\end{equation}
when {\bf l} and {\bf m} belong to the same sublattice and
\begin{equation}
C_{\bf l-m}=-\biggl\{\frac{2}{N}\sum_{\bf k}\!
 \stackrel{\vphantom{a}}{\vphantom{a}}^\prime\!\!
 \frac{{\rm e}^{i\bf k(m-l)}}{\sqrt{1-\eta^2\gamma^2_{\bf k}}}
 \biggl[\langle b_{\bf
 -k,-\sigma}b_{\bf k\sigma}\rangle_U-\eta\gamma_{\bf k}\biggl(\langle
 b^\dagger_{\bf k\sigma}b_{\bf k\sigma}\rangle_U+\frac{1}{2}\biggr)
 \biggr]\biggr\}^2
\end{equation}
\end{mathletters}
when {\bf l} and {\bf m} are on different sublattices.  In the
considered region of the $T$-$x$ plane the decay of spin
correlations (\ref{sc}) with the distance $|{\bf l-m}|$ is
nonexponential (see Fig.~9) which may be partly connected with
finite-size effects.  To estimate the spin correlation length we used
the formula
$$\xi^2=\sum_{\bf l}|{\bf l}|^2{\rm e}^{i\bf\Pi l}\langle
s^z_{\bf l}s^z_{\bf 0}\rangle /\bigl(2\sum_{\bf l}{\rm e}^{i\bf\Pi l}
\langle s^z_{\bf l}s^z_{\bf 0}\rangle\bigr).$$
For $T=116$~K the product $\xi\sqrt{x}$ is nearly constant and
approximately equal to the lattice spacing in the range $0.017\lesssim
x \lesssim 0.09$, in agreement with experiment in
La$_{2-x}$Sr$_x$CuO$_4$ \hfil \cite{birgeneau}.  \hfill For \hfil
larger \hfil $x$ \hfil at \hfil this \hfil temperature, \hfil $\xi$
\hfil becomes \hfil of \hfil the \hfil order \hfil of \hfil the

\hbox{
\psfig{figure=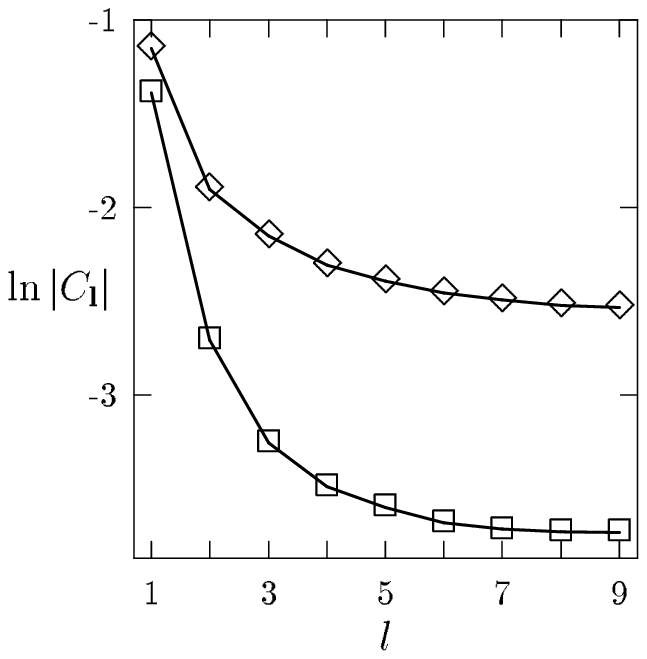}
\hspace{-2mm}
\raisebox{4cm}{\parbox{7.7cm}{\small Fig.~9.~The spin correlation
function along the lattice axis $[{\bf l}=(l,0)]$ for $x=0.027$
($\Diamond$) and $x=0.1$ ($\Box$) at $T=0$.}}}

\hbox{
\hspace*{2mm}
\psfig{figure=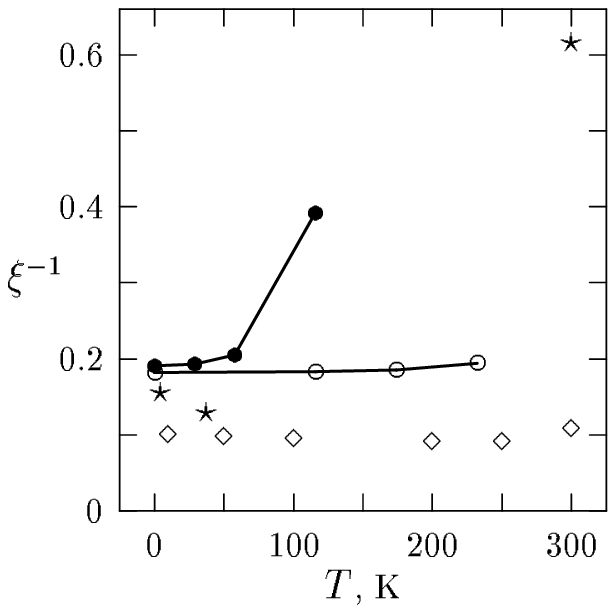}
\hspace*{3mm}
\raisebox{3.5cm}{\parbox{7.7cm}{\small Fig.~10.~The inverse
correlation length in units of the inverse lattice spacing for
$x\approx 0.03$ (open circles) and $x\approx 0.1$ (filled circles).
Values obtained from the neutron-scattering experiments in
La$_{1.86}$Sr$_{0.14}$CuO$_4$ \cite{zha} and in
La$_{1.96}$Sr$_{0.04}$CuO$_4$ \cite{birgeneau} are shown by stars and
diamonds, respectively.}}}

\noindent lattice spacing.  As mentioned, the use of the spin-wave
approximation (\ref{swa}) assumes that $\xi\gg a$.  We hope, however,
that our results obtained for the case $\xi\approx a$ (the region near
the curve in Fig.~1 for $x>0.09$) may give at least qualitatively a
correct description of this region.  When $x\lesssim 0.06$, our
calculated $\xi$ is nearly independent of temperature in the
considered range of $T$, as seen in Fig.~10 for the case $x\approx
0.03$ (for fixed $\mu$ the hole concentration is somewhat changed
with $T$; values of $x$ given in the figure captions here and below
are mean values for the considered temperature ranges).  This
behaviour agrees with experimental data in
La$_{1.96}$Sr$_{0.04}$CuO$_4$ \cite{birgeneau}, also shown in
Fig.~10.  For larger $x$ values $\xi^{-1}$ are also weakly
temperature-dependent at low $T$ and increase more rapidly as the
boundary in Fig.~1 is approached (see the data for $x\approx 0.1$ in
Fig.~10).  This behaviour is also in agreement with experimental
results in La$_{1.86}$Sr$_{0.14}$CuO$_4$ by G.~Aeppli et al.\ reported
in Ref.~\cite{zha} (stars in Fig.~10).  Notice, however, that the
growth of $\xi^{-1}$ predicted by the theory is somewhat more rapid
than that observed in experiment.  The saturation of $\xi$ with
decreasing $T$, demonstrated by Fig.~10, is the distinctive property
of the quantum disordered regime \cite{barzykin,chakravarty} in which
the system resides in the considered region of the $T$-$x$ plane.

In \hfil this \hfil region \hfil the \hfil hole \hfil contribution
\hfil $\chi^z_h$ \hfil to \hfil the \hfil susceptibility \hfil is
\hfil negligibly \hfil small \hfil in

\centerline{\psfig{figure=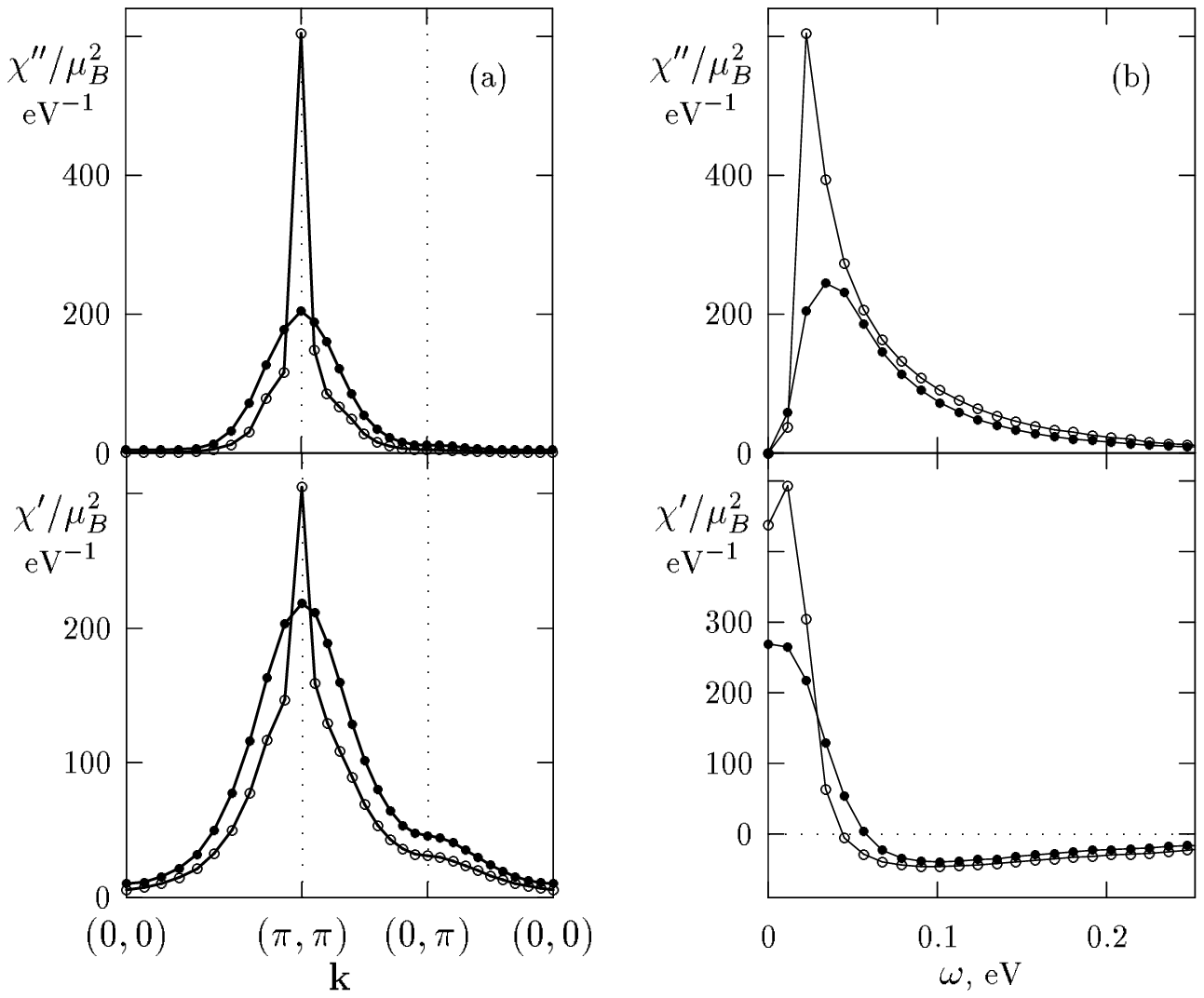}}

\noindent{\small Fig.~11.~Imaginary $\chi''$ and real $\chi'$ parts of
the magnetic susceptibility as functions of wave vector along the
symmetry lines in the Brillouin zone for $\omega\approx 0.022$~eV (a)
and frequency for ${\bf k}=(\pi,\pi)$ (b).  $x\approx 0.11$.  Open and
filled circles correspond to $T=58$~K and 116~K, respectively.}

\vspace{2ex}
\noindent comparison with the spin contribution $\chi^z_m$.  As seen
in Fig.~11, the magnetic susceptibility (\ref{chi}) is strongly peaked
around $(\pi,\pi)$.  This demonstrates strong antiferromagnetic
fluctuations which persist even in the case of comparatively short
correlation lengths of the order of the lattice spacing.  As known,
low-frequency incommensurate spin fluctuations are observed in the
normal state of Sr-doped La$_2$CuO$_4$ \cite{aeppli}.  This becomes
apparent in a four-peaked structure of ${\rm Im}\,\chi({\bf
q}\omega)$, the peaks being displaced from the commensurate position
to the points ${\bf q}_i=(\pi,\pi\pm\delta)$, $(\pi\pm\delta,\pi)$.
To investigate whether this incommensurability is connected with the
hole-magnon interaction in CuO$_2$ planes a more sophisticated
spin-wave approximation than that given in Eq.~(\ref{swa}) is needed.
We do not consider this point in the present paper.  Notice that the
frequency dependence of ${\rm Im}\,\chi({\bf q}\omega)$ in Fig.~11b is
close to that observed in normal-state La$_{1.86}$Sr$_{0.14}$CuO$_4$
at wave vectors ${\bf q}_i$ \cite{zha}.  Both the total peak intensity
at these wave vectors, the position of the peak, and its temperature
dependence are close to those shown in this figure.

We used the obtained magnetic susceptibility for calculating the
spin-lattice relaxation times at the Cu and O sites $^{67}T_1$ and
$^{17}T_1$ and the Cu spin-echo decay time $T_{2G}$ by using the
equations \cite{barzykin}
\begin{eqnarray}
&&\frac{1}{^\alpha T_1T}=\frac{1}{2\mu^2_BN}\sum_{\bf q}
 \,\mbox{}^\alpha\!
 F({\bf q})\frac{{\rm Im}\,\chi({\bf q}\omega)}{\omega},
 \quad\omega\rightarrow 0,\nonumber\\[0.4ex]
&&\frac{1}{T_{2G}}=\sqrt{\frac{0.69}{128}}
 \left(^{63}\gamma_n\right)^2\Biggl\{
 \frac{1}{N}\sum_{\bf q}F_e^2({\bf q})\left[{\rm Re}\,\chi({\bf q}0)
 \right]^2-\left[\frac{1}{N}\sum_{\bf q}F_e({\bf q})
 {\rm Re}\,\chi({\bf q}0)\right]^2\Biggr\}^{1/2},\label{nmr}\\[1ex]
&&^{63}\!F({\bf q})=\left[A_\bot+4B\gamma_{\bf q}\right]^2,\quad
 F_e({\bf q})=\left[A_\|+4B\gamma_{\bf q}\right]^2,\quad
 ^{17}\!F({\bf q})=2C^2\left[1+\cos(q_x)\right]^2,\nonumber
\end{eqnarray}
where $^{63}\gamma_n$ is the Cu nucleus gyromagnetic ratio, $B=3.82
\cdot 10^{-7}$~eV, $A_\bot=0.84B$, $A_\|=-4B$, and $C=0.91B$.  The
spin-lattice relaxation times in Eq.~(\ref{nmr}) correspond to the
applied static magnetic field perpendicular to the CuO$_2$ plane.
Results are shown in Fig.~12 together with the calculated
$\chi^0=\chi(0,0)$ and the respective experimental results obtained in
YBa$_2$Cu$_3$O$_{6.63}$ \cite{barzykin,takigawa91,takigawa94}. From our
results for different hole concentrations we selected for this figure
those which appeared to be closest to the experimental data.

\vspace{1ex}
\hspace*{-10mm}\psfig{figure=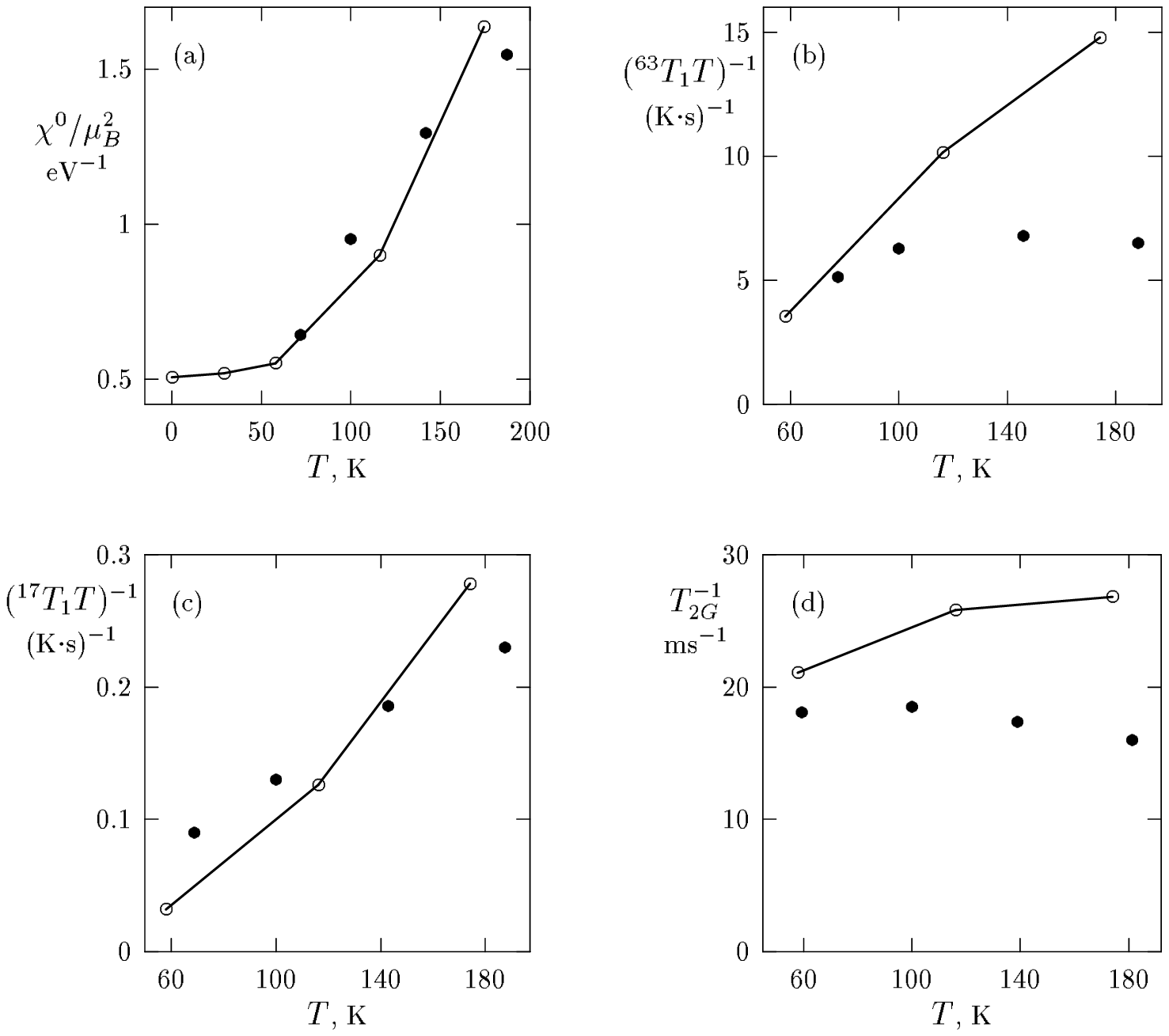}

\noindent{\small Fig.~12.~The static spin susceptibility (a),
$1/(T_1T)$ at the Cu (b) and O (c) sites for ${\bf H} || {\bf c}$,
and $1/T_{2G}$ (d) calculated for $x\approx 0.05$ (open circles).  The
experimental data for YBa$_2$Cu$_3$O$_{6.63}$ are shown by filled
circles.  $\chi^0$ was inferred from the Knight shift data
\cite{barzykin}, $1/(T_1T)$ and $1/T_{2G}$ are from
Refs.~\cite{takigawa91} and \cite{takigawa94}, respectively.}

\noindent Those results are for $x\approx 0.05$ which is apparently
somewhat smaller than the hole concentration in
YBa$_2$Cu$_3$O$_{6.63}$.  With increasing $x$ our calculated $\chi^0$
and $(^\alpha T_1T)^{-1}$ increase, remaining of the same order of
magnitude as those shown in Fig.~12, while $T^{-1}_{2G}$ slightly
decreases.  These concentration dependences of $\chi^0$ and
$T^{-1}_{2G}$ agree with experimental observations
\cite{barzykin,takigawa94,walstedt,imai}, though for $\chi^0$ the
theoretical dependence is stronger than that observed experimentally.
The situation is apparently more difficult for $(^{63}T_1T)^{-1}$.  In
underdoped YBa$_2$Cu$_3$O$_{6+y}$ this quantity decreases with doping
\cite{berthier}, while in HgBa$_2$CuO$_{4+\delta}$, it increases
\cite{yasuoka}.  Apparently a more elaborate model is necessary for
the description of this concentration dependence.  Analyzing Fig.~12
we conclude that the $t$-$J$ model is able to describe correctly the
temperature dependences of the depicted quantities and to give their
proper orders of magnitude in the quantum disordered regime.

As seen in Fig.~12, $(T_1T)^{-1}$ and $\chi^0$ decrease with
decreasing temperature.  Such behaviour observed for $T>T_c$ in
underdoped cuprates \cite{imai89,rossat} is considered as an indication
of the pseudogap in the spectrum of magnetic excitations
\cite{imai89,rossat,sokol,barzykin}. Our results corroborate this point
of view --- analyzing Eq.~(\ref{chi}) and our numerical data we came to
the conclusion that the mentioned temperature dependence is mainly
connected with the occupation of low-frequency magnons, rather than the
temperature variation of the magnon spectral intensities.  Due to the
pseudogap this occupation decreases with temperature.

\section{Summary}
In this work we applied the modified spin-wave theory with the
additional constraint of zero staggered magnetization to the
two-dimensional $t$-$J$ model in the paramagnetic state.  In the Born
approximation the constraint equation (\ref{cond})
and the self-energy equations (\ref{se}) for
the hole and magnon Green's functions form the self-consistent set which
was solved numerically for finite hole concentrations $x$ and
temperatures $T$.  The constraint can be fulfilled in the region of the
$T$-$x$ plane below the curve in Fig.~1.

A number of unusual features of photoemission spectra in cuprate
perovskites are satisfactorily reproduced by the obtained results.
Among these features are the general shape of the electron energy
spectrum and its evolution from the narrow spin-polaron band at small
doping to the much wider band for moderate doping.  Both by the energy
position and by the extension in the Brillouin zone our calculated
extended saddle point reproduces well the van Hove singularity of
photoemission spectra.  Also the obtained Fermi surface with hidden and
two-dimensional parts is close to that observed experimentally.  In the
calculated Fermi surface the size of the hole pockets varies only
slightly with the growth of the hole concentration.  This growth is
mainly connected with an increase of the quasiparticle weights of states
occupied by holes.  In our calculated hole spectrum the pseudogap has
the magnitude, symmetry and the concentration dependence which are
similar to those observed in photoemission.  This pseudogap is not
connected with superconducting fluctuations which were not included in
the calculations.  It arises due to the specific dispersion of the
spin-polaron band a part of which is retained near the Fermi level at
moderate doping and gives the most intensive maxima in the spectral
function.  The persistence of this band is an indication of strong
electron correlations retained at moderate doping.

We calculated also a number of magnetic characteristics of the $t$-$J$
model.  We found that in the region shown in Fig.~1 the
temperature variation of the spin correlation length is typical for the
quantum disordered regime and that this dependence as well as the
concentration dependence of the correlation length are close to those
observed in cuprates.  The magnon spectrum contains the pseudogap which
manifests itself in the temperature dependence of the static spin
susceptibility and spin-lattice relaxation rates at the Cu and O sites.
Our calculated values of these quantities and the Cu spin-echo decay
rate, their temperature and concentration dependences are in qualitative
and in some cases in quantitative agreement with experiment.

The considered phase with the above-discussed properties differs
essentially from the conventional metal.  As mentioned, this phase
exists in the region bounded by the curve in Fig.~1.  Outside of
this region for $x\gtrsim 0.12$ we found the energy spectrum of the
usual metal.

\acknowledgements
This work was partially supported by the ESF grant No.~2688 and by the
WTZ grant (Project EST112.1) of the BMBF.

\end{document}